\documentclass[epj,nopacs,onecolumn,fleqn]{svjour}          

\RequirePackage[T1]{fontenc}

\smartqed  

\RequirePackage{graphicx}
\RequirePackage{mathptmx}      
\RequirePackage{flushend}
\RequirePackage[numbers,sort&compress]{natbib}
\RequirePackage[colorlinks,citecolor=blue,urlcolor=blue,linkcolor=blue]{hyperref}

\usepackage[T1]{fontenc}
\usepackage{amsmath}    
\usepackage{amssymb}    %
\usepackage{hyperref}
\usepackage{float}	

\newcommand{\ttt}{t\bar{t}}
\newcommand{\alphas}{\alpha_{s}}

\newcommand{\gev}{\,\textrm{GeV}}

\newcommand{\ifb}{\,\textrm{fb}^{-1}}
\newcommand{\tev}{\,\textrm{TeV}}
\newcommand{\TO}{\rightarrow}

\newcommand{\ord}{\mathcal{O}}

\def\beq{\begin{equation}}
\def\beqn{\begin{eqnarray}}
\def\eeq{\end{equation}}
\def\eeqn{\end{eqnarray}}
\def\beal{\begin{align}}
\def\endal{\end{align}}

\newcommand\aNLO{{\sc\small MadGraph5\_aMC@NLO}}

\numberwithin{equation}{section}

\journalname{Eur. Phys. J. C}

\begin{document}

\title{
The impact of the photon PDF and electroweak corrections on $t \bar t$ distributions
}
\author{D.~Pagani\inst{1}
        \and
        I.~Tsinikos\inst{1} 
        \and
        M.~Zaro\inst{2,3} 
}

\institute{Centre for Cosmology, Particle Physics and Phenomenology (CP3), Universit\'e Catholique de 
            Louvain, Chemin du Cyclotron 2, B-1348 Louvain-la-Neuve, Belgium
          \and
          Sorbonne Universit\'es, UPMC Univ. Paris 06, UMR 7589, LPTHE, F-75005, Paris, France
          \and
          CNRS, UMR 7589, LPTHE, F-75005, Paris, France
}

\abstract{We discuss the impact of EW corrections on differential distributions in top-quark pair production at the LHC and future hadron colliders, 
focusing on the effects of initial-state photons. Performing a calculation at Next-to-Leading Order QCD+EW accuracy, we investigate in 
detail the impact of photon-initiated channels on central values as well as PDF and scale uncertainties, both at order $\alpha_s\alpha$ 
and $\alpha_s^2\alpha$. We present
predictions at 13 and 100 TeV, and provide results for the 8 TeV differential measurements performed by ATLAS and CMS. 
A thorough comparison of results obtained with the NNPDF2.3QED and CT14QED PDF sets is performed. While contributions
due to the photon PDF are negligible with CT14QED, this is not the case for NNPDF2.3QED, where such contributions are sizeable and show 
large PDF uncertainties. On the one hand, we show that differential observables in top-pair production, in particular top-quark and 
$t\bar t$ rapidities, can be used to improve the determination of the photon PDF within the NNPDF approach. On the other hand, with current 
PDF sets, we demonstrate the necessity of including EW corrections and photon-induced contributions for a correct determination of both the 
central value and the uncertainties of theoretical predictions.}

\date{\today}

\maketitle

\vspace*{-12cm}
\noindent
{\small{CP3-16-28}}
\vspace*{10.4cm}

\section{Introduction}
\label{sec:intro}

The production of a pair of top quarks ($\ttt$) is one of the key processes in the physics program at the LHC. On the one hand, this process 
is an ideal testing ground for strong interactions, which can be probed in the perturbative regime, and more in general for the Standard Model (SM) interactions.  On the other hand, because the experimental signatures emerging from $\ttt$ production can feature leptons, missing transverse energy and ($b$)-jets, this process enters as a background in several experimental analyses and searches, notably those related to the Higgs boson and  new physics.
The study of $\ttt$ production gives also access to many of the top-quark properties and can be exploited for the determination of the Parton-Distribution-Functions (PDF) of the proton~\cite{Beneke:2012wb,Czakon:2013tha,Guzzi:2014wia}. Furthermore, since the top quark is the heaviest particle in the SM, it is often a proxy to new physics in Beyond-the-Standard-Model (BSM) scenarios. Thus,  $\ttt$ production can be exploited to precisely test  SM predictions as well as to search for or put constraints on BSM interactions.

For these reasons, a proper modelling of the $\ttt$ signal and a sound understanding of all effects that can lead to distortions in the distributions are mandatory. This explains the great efforts which have been put in place in order to improve the accuracy of theoretical predictions: 
 inclusive and differential cross-sections have been calculated up to Next-to-Next-to-Leading-Order (NNLO) accuracy in QCD~\cite{Czakon:2013goa, Czakon:2015owf}, 
and including soft gluon effects up to Next-to-Next-to-Leading-\-Loga\-rith\-mic (NN\-LL) 
accuracy~\cite{Moch:2008qy, Kidonakis:2009ev, Czakon:2009zw, Ahrens:2010zv, Beneke:2010da, Kidonakis:2010dk, Ahrens:2011mw, Ahrens:2011px, Beneke:2011mq, Cacciari:2011hy,  Czakon:2013tha, Pecjak:2016nee}. Notably, higher-order QCD corrections lead to the reduction of scale uncertainties down to very few percents. At this level of accuracy effects due to Electro-Weak (EW) corrections cannot be neglected, especially for differential distributions, where the expected hierarchy between 
the strong and electroweak interactions ($\alpha_s \gg \alpha$) may not be respected.  On the one hand, when large scales are probed ($Q \gg m_W$), the so-called Sudakov logarithms render EW corrections large and negative. On the other hand, EW corrections receive also contributions from channels
that feature photons in the initial state,  thus they depend on the PDF of the photon. As we will show in this work, the impact of photons in the initial state strongly depends on the PDF set employed and can give a positive contribution of non-negligible size, leading to possible compensations of the corrections induced by Sudakov logarithms.

While weak ~\cite{Beenakker:1993yr,Kuhn:2005it, Bernreuther:2005is, Kuhn:2006vh,Bernreuther:2006vg,Bernreuther:2008md, Manohar:2012rs, Kuhn:2013zoa,Campbell:2015vua}, QED~\cite{Hollik:2007sw} and EW (weak+QED) \cite{Bernreuther:2010ny, Hollik:2011ps, Kuhn:2011ri, Bernreuther:2012sx} corrections to $\ttt$ production at hadron colliders have been known for quite some time and the effect of Sudakov logarithms  has been quantified and studied, the impact of photon-induced subprocesses has been addressed only in \cite{Hollik:2007sw} for only the $g\gamma \TO \ttt$ channel at LO and using the only PDF set with the photon distribution available at that time, {\it i.e.}, {\sc\small  MRST\-2004\-QED} \cite{Martin:2004dh}. 

In this work we (re-)evaluate, besides the NLO QCD corrections, the complete set of NLO EW corrections to top-quark pair production and carefully assess their impact on differential distributions for proton--proton collisions at 8, 13 and 100 TeV centre-of-mass energy. For the first time, photon-induced contributions are taken into account both at $\ord(\alphas \alpha)$, from the $g\gamma$-channel at tree-level, and at $\ord(\alphas^2 \alpha)$, from the $g\gamma$ as well as the $q\gamma(\bar{q}\gamma)$ initial states arising in the NLO EW corrections.  For all the results presented here, the calculation has been performed within the {\aNLO} framework~\cite{Alwall:2014hca}, thanks to an extension of the code that allows to automatically calculate NLO QCD and EW corrections \cite{Frixione:2014qaa, Frixione:2015zaa}.
In order to have a reliable estimate of the photon-induced contribution and of its uncertainties, we evaluate the impact of electroweak corrections with both the {\sc\small NNPDF\-2.3\-QED}~\cite{Ball:2013hta} and {\sc\small CT14QED}~\cite{Schmidt:2015zda} PDF sets, the only modern sets that include LO QED contribution in the DGLAP evolution and, with very different assumptions and strategies in the extraction from data, the photon distribution~\footnote{We acknowledge that two new studies on the determination of the photon PDF~\cite{Manohar:2016nzj, Harland-Lang:2016kog} have appeared during the publication of our work.}.

We show that, besides higher-order QCD corrections, it is also necessary to take into account  EW corrections, including photon-induced contributions, for a correct determination of both the central value and the uncertainty of theoretical predictions. This is particularly important in the context of the  ongoing discussion on NLO-accurate event generators and the compatibility with experimental data for the $p_T(t)$ distribution at the LHC 8-TeV measurements \cite{CMS:2015tpz} and in view of the measurements at 13 TeV.
Furthermore, we show that differential observables in top-pair production, in particular top-quark and $t \bar t$ rapidities, can be used to improve the determination of the photon PDF within the {\sc\small NNPDF} approach, while in the {\sc\small CT14QED} approach $t \bar t$ production is not sensitive to the photon-induced contributions.

The structure of our paper is the following: in sect.~\ref{sec:calc} we present the framework employed to perform our calculation and discuss relevant input parameters. In sect.~\ref{sec:PDF} we discuss differences among existing PDF sets which provide a photon distribution ({\sc\small NNPDF\-2.3\-QED}, {\sc\small CT14QED} and the older {\sc\small MRST\-2004\-QED} set). We describe the different theoretical approaches employed in the sets, and we compare central values and uncertainties for the photon PDF  and the parton luminosities relevant for our calculation. In sect.~\ref{sec:res} we show predictions at 13 and 100 TeV, and  we compare results with and without the contribution of photon-induced processes and using the {\sc\small NNPDF\-2.3\-QED} or the {\sc\small CT14QED} PDF set.  In sect.~\ref{sec:data} we analyse the impact of EW corrections and the photon PDF for specific measurements performed by ATLAS and CMS at 8 TeV. We compare the usage of {\sc\small NNPDF\-2.3\-QED} with a standard modern set with QCD-only partons and DGLAP evolution.  We give our conclusions and outlook in sect.~\ref{sec:concl}.
\section{Calculation setup and input parameters}
\label{sec:calc}

The calculation has been performed in a completely automated way and we do not describe here the technical details; they can be found in \cite{Frixione:2014qaa, Frixione:2015zaa}, where the $\ttt H$ process has been calculated in the same framework. In the following, we only want to match the notation of this paper to the one introduced in  \cite{Frixione:2014qaa} and precisely define the quantities included in our calculation, specifying those that depend on the photon PDF. 

In the case of $pp \rightarrow \ttt$ process a generic observable $\Sigma^{\ttt}$  can be expanded simultaneously in the QCD and EW coupling constants as:
\noindent
\begin{equation}
\Sigma^{\ttt}(\alpha_s,\alpha) = \sum_{m+n\geq 2} \alpha_s^m \alpha^n \Sigma_{m+n,n}\, .
\end{equation}
\noindent
Following this notation, one can separate the LO ($m+n=2$) and NLO ($m+n=3$) contributions as follows:
\begin{align}
\Sigma^{\ttt}_{\rm LO}(\alpha_s,\alpha) &= \alpha_s^2 \Sigma_{2,0} + \alpha_s \alpha \Sigma_{2,1} + \alpha^2 \Sigma_{2,2} \equiv \nonumber\\
&\equiv\Sigma_{\rm LO,1} + \Sigma_{\rm LO,2} + \Sigma_{\rm LO,3}\, ,  \nonumber\\
\Sigma^{\ttt}_{\rm NLO}(\alpha_s,\alpha) &= \alpha_s^3 \Sigma_{3,0} + \alpha_s^2 \alpha \Sigma_{3,1} + \alpha_s \alpha^2 \Sigma_{3,2} + \alpha^3 \Sigma_{3,3} \equiv 
\nonumber\\
&\equiv \Sigma_{\rm NLO,1} + \Sigma_{\rm NLO,2} + \Sigma_{\rm NLO,3} + \Sigma_{\rm NLO,4} \;.
\label{eq:blobs}
\end{align}
\noindent
In our results we include the $\Sigma_{\rm LO,1}$, $\Sigma_{\rm LO,2}$, $\Sigma_{\rm NLO,1}$ and $\Sigma_{\rm NLO,2}$  terms.
We checked that the remaining terms are subleading as expected, giving results of the order or below $1\%$ of the LO,1 contribution both in the total cross section and at the differential level. 
 In order to help the reader we further define the quantities
\noindent
\begin{align}
\Sigma_{\rm LO~QCD} &\equiv \Sigma_{\rm LO,1} \, , &    \Sigma_{\rm NLO~QCD} &\equiv  \Sigma_{\rm NLO,1}\, , \label{blobbegin} \\
\Sigma_{\rm LO~EW} &\equiv \Sigma_{\rm LO,2} \, ,   &     \Sigma_{\rm NLO~EW} &\equiv  \Sigma_{\rm NLO,2}\, ,
\end{align}
\begin{eqnarray}
 \Sigma_{\rm QCD} &\equiv& \Sigma_{\rm  LO~QCD} + \Sigma_{\rm NLO~QCD} \,, \\ 
\Sigma_{\rm EW} &\equiv& \Sigma_{\rm LO~EW} + \Sigma_{\rm NLO~EW}\,,\\
    \Sigma_{\rm QCD+EW} &\equiv& \Sigma_{\rm  QCD} + \Sigma_{\rm EW} \label{blobend} \,.
\end{eqnarray}
\noindent

In the following text with the term ``EW corrections'' we will in general refer to the quantity $\Sigma_{\rm EW}$, while we will use ``NLO EW corrections'' for $\Sigma_{\rm NLO~EW}$.  
At variance with refs.~\cite{Frixione:2014qaa, Frixione:2015zaa}, in our calculation we do not consider the effect due to the Heavy-Boson-Radiation (HBR). Although the LO cross sections of $pp\TO \ttt V$, $V=W^\pm,Z,H$ processes may in principle contribute at the same perturbative order of $\Sigma_{\rm NLO~EW}$ to the inclusive $\ttt$ production, in this work we assume that $\ttt$ and $\ttt V$ final states can always be distinguished.

The quantities that depend on the photon PDF are $\Sigma_{\rm LO~EW} $ and $\Sigma_{\rm NLO~EW}$. The $g \gamma  \TO \ttt$ process contributes to $\Sigma_{\rm LO~EW} $ and to $\Sigma_{\rm NLO~EW}$. In addition, $\Sigma_{\rm NLO~EW}$ receives contributions from the $q\gamma\TO\ttt q$ and $\bar{q}\gamma\TO\ttt \bar{q}$ processes at the tree level, which feature initial-state singularities that have to be subtracted. As in the case of $q\bar{q}\TO \ttt \gamma$ processes, the subtracted QED singularities are taken into account in the DGLAP evolution in {\sc\small  MRST\-2004\-QED}, {\sc\small NNPDF\-2.3\-QED} and {\sc\small CT14QED}. Thus, for theoretical consistency, these three PDF sets should be preferred whenever NLO EW corrections are computed.
 However, since the QED accuracy of DGLAP evolution is only  LO in all the three PDF sets, one could not technically claim at the moment NLO QED and thus NLO EW accuracy for hadronic predictions. Even before considering the numerical results in sects.~\ref{sec:res} and \ref{sec:data}, this issue points to the necessity of a better determination of the photon PDF by means of fits which include NLO QED effects in the DGLAP evolution \cite{deFlorian:2015ujt, deFlorian:2016gvk}.\\

Our calculation is performed using the following input parameters
\noindent
\begin{eqnarray}
m_t = 173.3 \text{ GeV}\, , \quad m_H = 125.09 \text{ GeV} \, , \nonumber\\
 \quad m_W = 80.385 \text{ GeV} \, , \quad m_Z = 91.1876 \text{ GeV} \,  ,
\end{eqnarray}
\noindent
and setting all the other fermion masses to zero. All masses are renormalised on-shell and all decay widths are set to zero. 
The renormalisation of $\alphas$ is performed in the five-flavour scheme, while EW parameters are chosen in the $G_\mu$-scheme, with
\begin{equation}
 G_\mu = 1.1663787 \cdot 10^{-5} \text{ GeV}^{-2} \,.
\end{equation}
Since NLO EW corrections of $\mathcal{O}(\alpha_{s}^{2}\alpha)$ to $\ttt$ hadroproduction do not involve the renormalisation of $\alpha$, the choice of a different EW scheme will not change our results in a visible way. The CKM matrix is taken as the identity.

Unless differently specified, we use a dynamical reference scale for the central values of the renormalisation ($\mu_r$) and factorisation ($\mu_f$) scales defined as 
\noindent
\begin{equation}
\mu = \frac{H_T}{2} = \frac{1}{2} \sum_{i} m_{T,i} \; ,
\end{equation}
\noindent
where the sum of the transverse masses runs over all the final-state particles. In all cases theoretical uncertainties due to missing higher orders are estimated via independent variations of $\mu_r$ and $\mu_f$ in the interval $\{\mu/2,2\mu\}$. 

It is worth to note that the {\sc\small NNPDF\-2.3\-QED} set is in the variable-flavour scheme with six active flavours, which for $\mu>m_{t}$ is equivalent to the six-flavour scheme. On the contrary,  in  {\sc\small CT14QED} the active flavours are five, leading to the five-flavour scheme also for $\mu>m_{t}$.  As we said, we renormalise $\alpha_s$ in the five-flavour scheme for all the predictions;  for the comparison  between the   {\sc\small NNPDF\-2.3\-QED}  and {\sc\small CT14QED} results we simply change the PDF set without modifying the calculation framework. The change of scheme can be easily performed by following the recipe described in \cite{Badger:2016bpw} and based on \cite{Cacciari:1998it}, which, at NLO QCD + NLO EW accuracy, has a  direct effect only on the $q\bar{q}$-induced  contribution to $ \Sigma_{\rm NLO~QCD}$. We explicitly verified that the numerical impact of such a change of renormalisation scheme is always much smaller of the scale uncertainty  and furthermore cancels in any ratio involving $ \Sigma_{\rm NLO~QCD}$ both at the numerator and the denominator. Thus, it has not any influence on the discussion presented in this work.

\section{Photon PDF and parton luminosities}
\label{sec:PDF}
\begin{figure*}[t]
\centering
\includegraphics[width=0.32\textwidth, clip=true, trim=0cm 8cm 4.5cm 3.5cm]{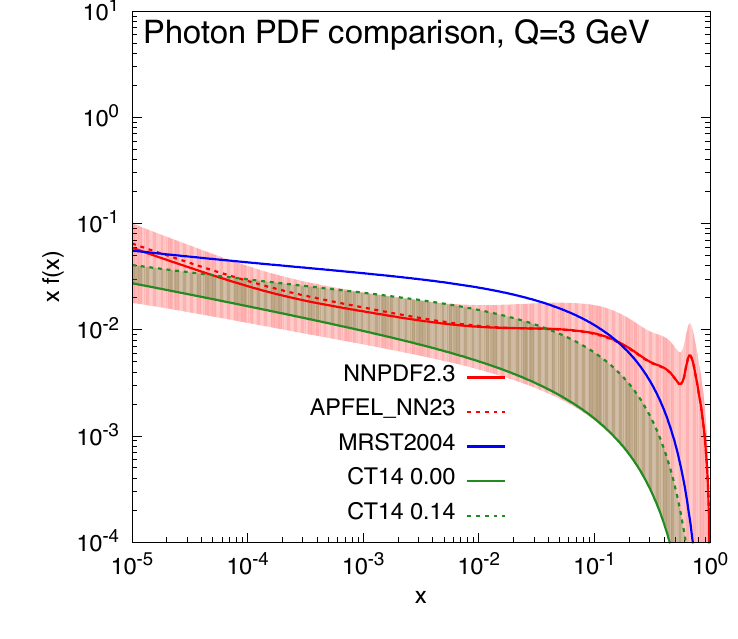}
\includegraphics[width=0.32\textwidth, clip=true, trim=0cm 8cm 4.5cm 3.5cm]{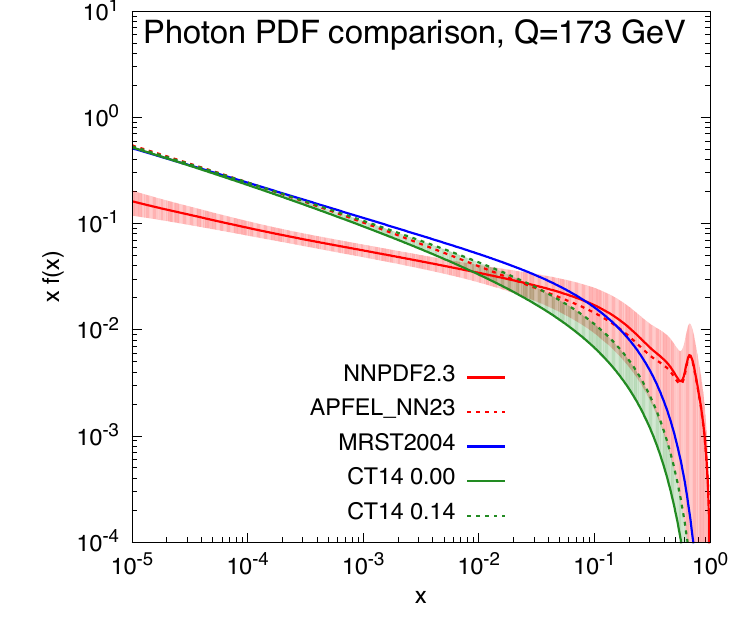}
\includegraphics[width=0.32\textwidth, clip=true, trim=0cm 8cm 4.5cm 3.5cm]{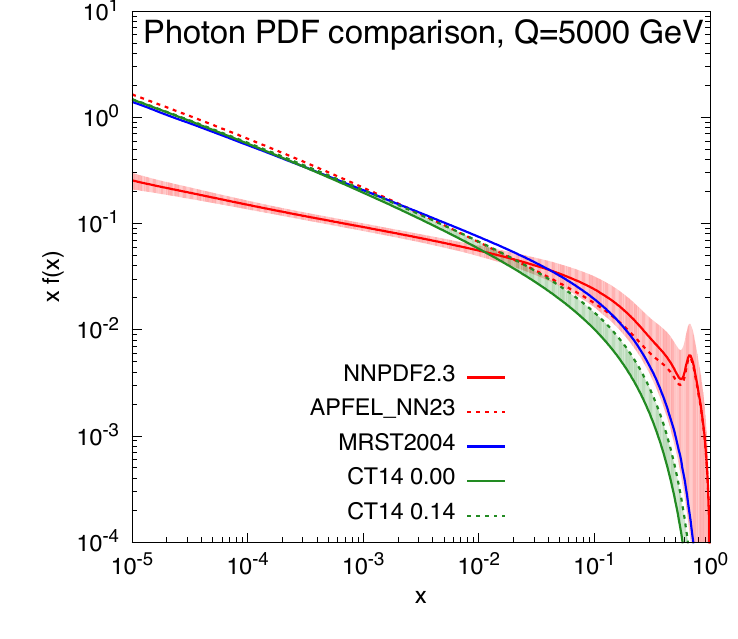}
\caption{Comparison of the photon PDF for {\sc\small MRST\-2004\-QED}, {\sc\small CT14QED}, {\sc\small NNPDF\-2.3\-QED} and {\sc\small APFEL\_NN23} sets at $Q=3~\gev$, $Q=173~\gev$ and $Q=5000~\gev$.}
\label{fig:photonPDF}
\end{figure*}

\begin{figure*}[t]
\centering
\includegraphics[width=0.32\textwidth, clip=true, trim=0cm 4.5cm 4.5cm 3.5cm]{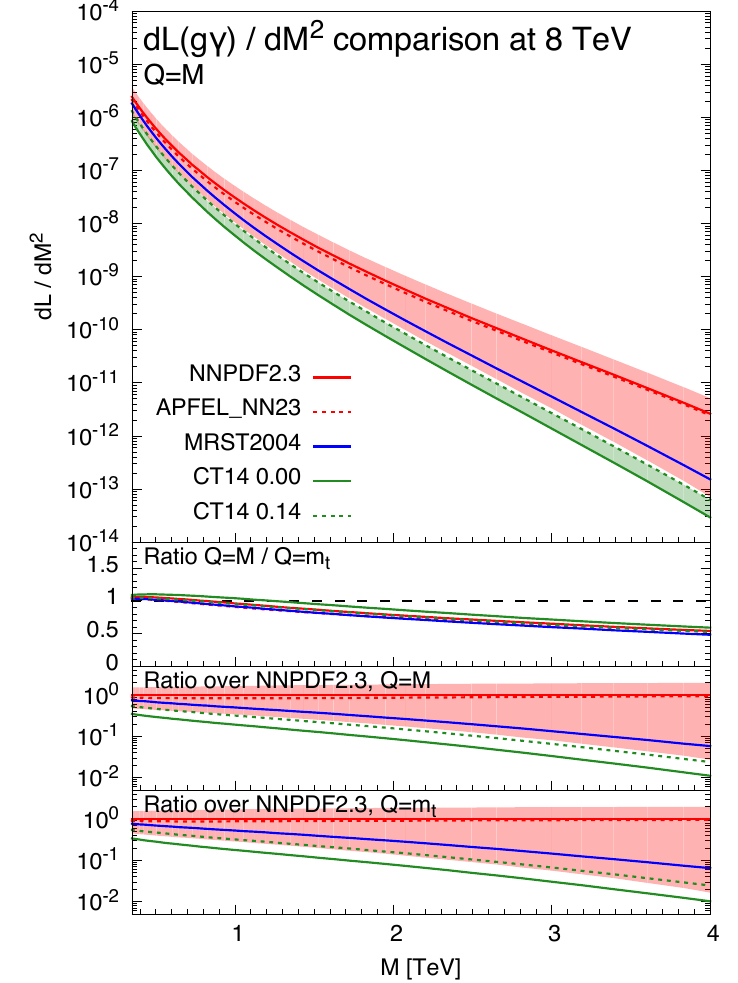}
\includegraphics[width=0.32\textwidth, clip=true, trim=0cm 4.5cm 4.5cm 3.5cm]{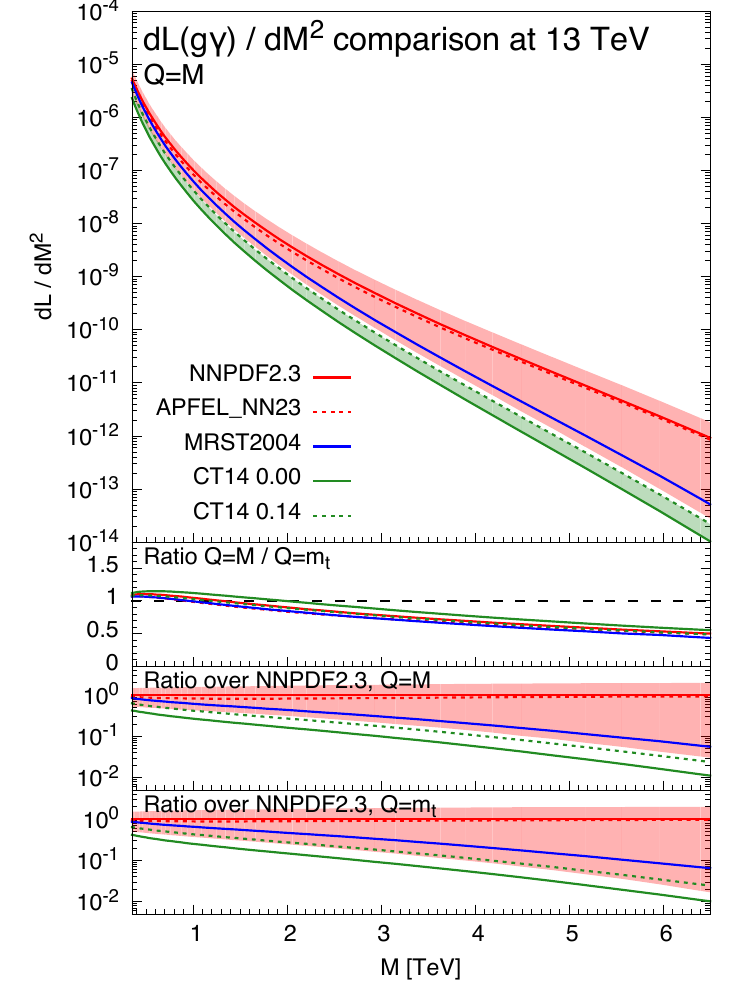}
\includegraphics[width=0.32\textwidth, clip=true, trim=0cm 4.5cm 4.5cm 3.5cm]{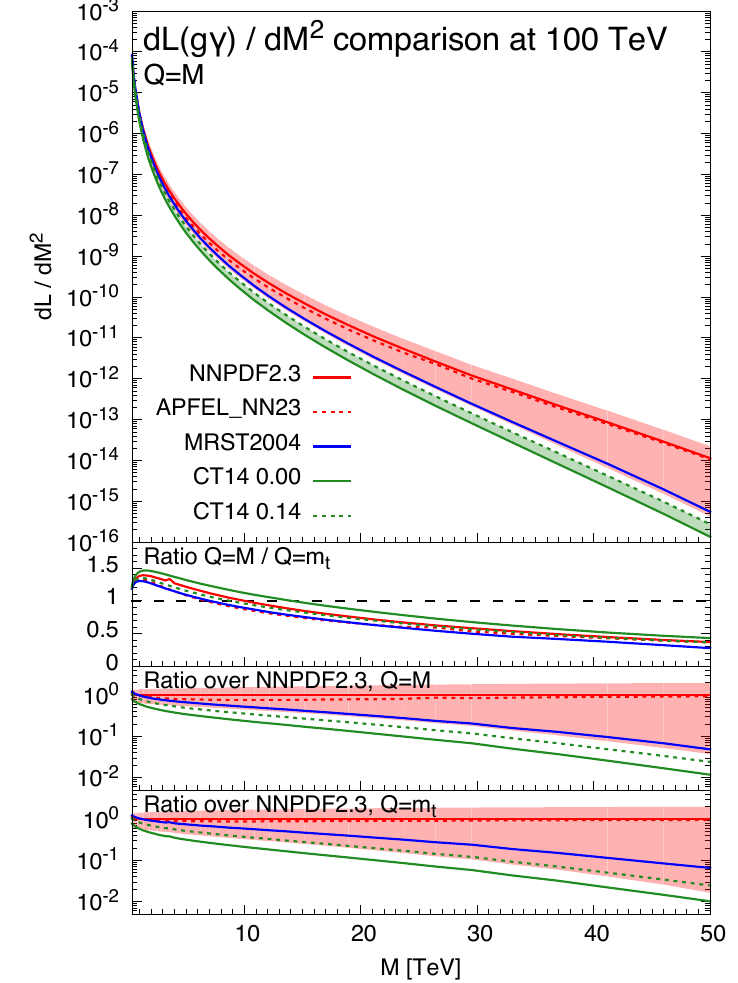}
\caption{Comparison of the $g \gamma$ luminosity for {\sc\small MRST\-2004\-QED}, {\sc\small CT14QED}, {\sc\small NNPDF\-2.3\-QED} and {\sc\small APFEL\_NN23} sets at 8, 13 and 100 TeV.}
\label{fig:aglumo}
\end{figure*}
\begin{figure*}[t]
\centering
\includegraphics[width=0.32\textwidth, clip=true, trim=0cm 4.5cm 4.5cm 3.5cm]{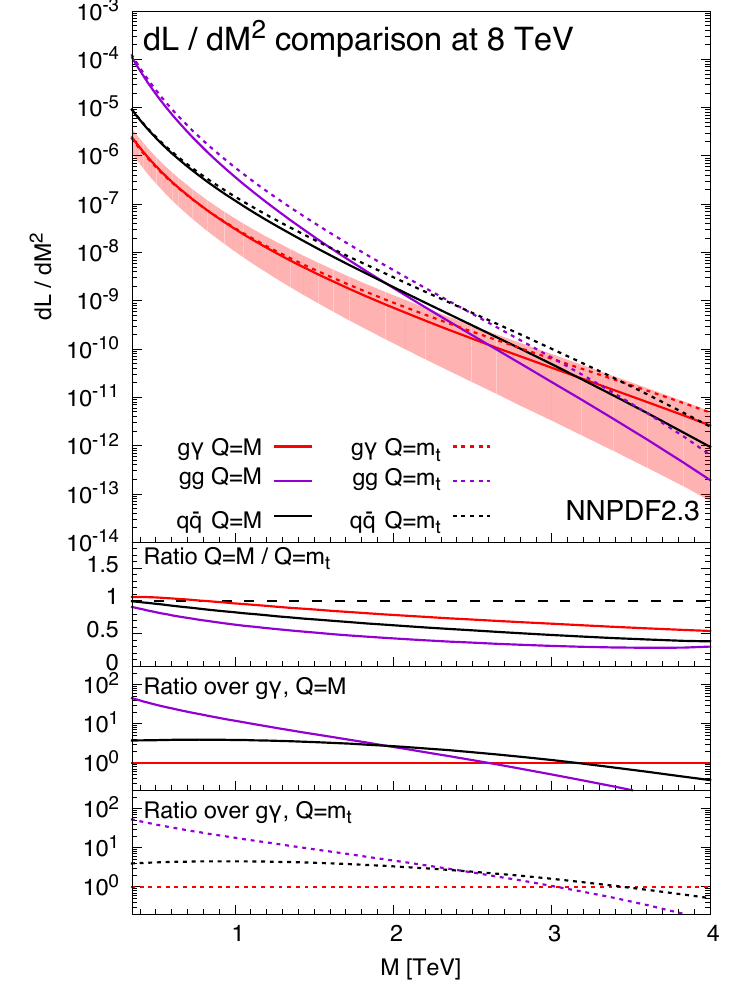}
\includegraphics[width=0.32\textwidth, clip=true, trim=0cm 4.5cm 4.5cm 3.5cm]{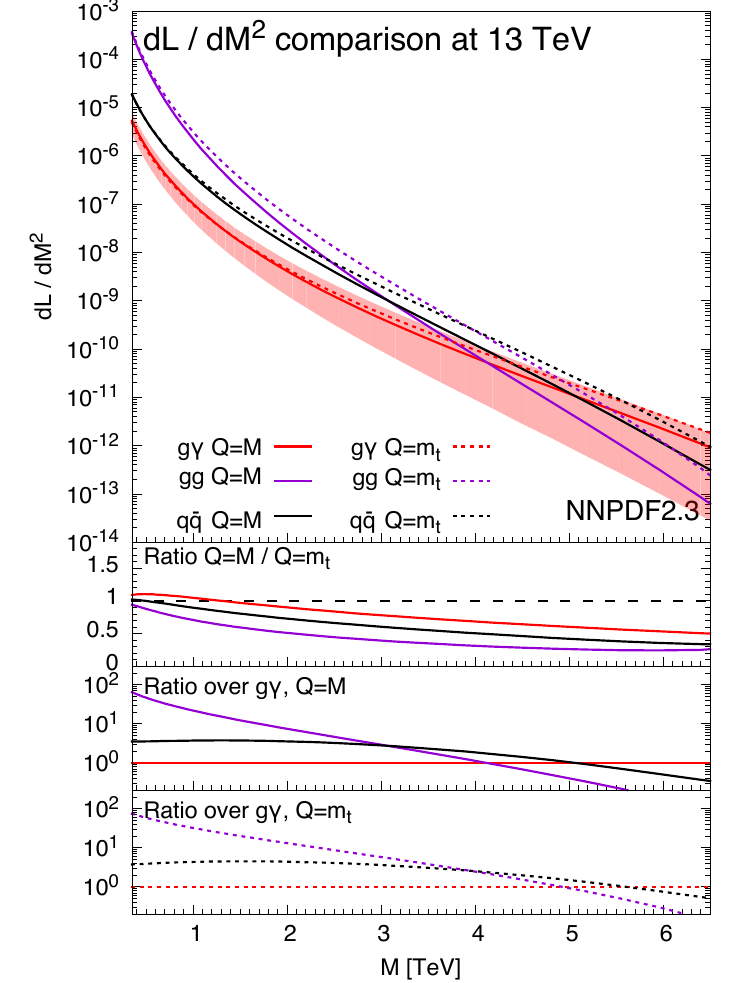}
\includegraphics[width=0.32\textwidth, clip=true, trim=0cm 4.5cm 4.5cm 3.5cm]{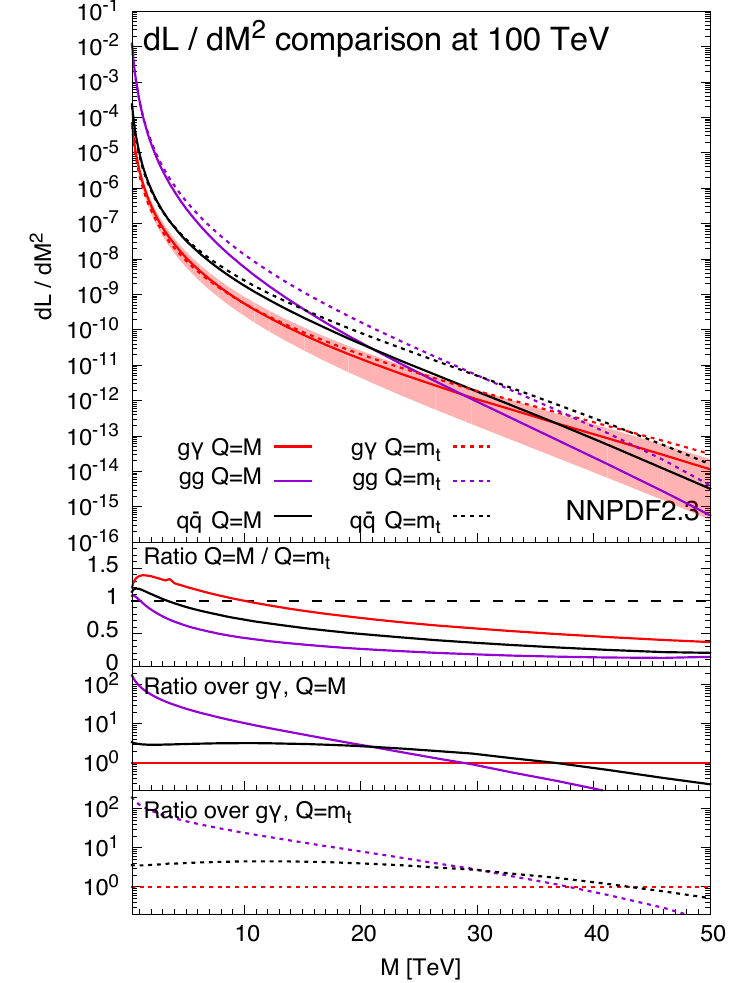}
\caption{Comparison of the $g \gamma$, $gg$ and $q \bar q$ luminosities for the {\sc\small NNPDF\-2.3\-QED} set at 8, 13 and 100 TeV.}
\label{fig:luminosities}
\end{figure*}

In this section  we discuss in some details the different modelling of the photon PDF in the {\sc\small NNPDF\-2.3\-QED}, {\sc\small CT14QED}\footnote{
    {\sc\small CT14QED} provides two kinds of sets, one with only the incoherent component of the photon PDF and another one with both the coherent and incoherent 
    components. In our work
    we have used the first
    set. We have checked that the inclusion of the coherent component in the photon PDF does not significantly alter our findings. The predictions
    obtained with the photon PDF with momentum fraction $p_0^\gamma=0.00\%$ including both components are very similar, 
    in the $x$ and $Q$ ranges relevant for our study, 
    to those from the incoherent-only photon PDF with $p_0^\gamma=0.14\%$.}
and {\sc\small MRST\-2004\-QED} sets.

Although for all the three PDF sets the DGLAP evolution is performed at NLO QCD + LO QED accuracy~\footnote{In the case of {\sc\small NNPDF\-2.3\-QED}, PDFs at NNLO QCD + LO QED accuracy as well as at LO QCD + LO QED accuracy are also available. However we considered here only the NLO QCD + LO QED case, consistently with the other PDF sets discussed. }, very different and crucial assumptions underlie the determination of the photon PDF $\gamma(x,Q)$.
These differences mainly concern the following three aspects. First, the ansatz for $\gamma(x,Q)$ at the initial scale $Q=Q_0$. Second, the different datasets which are used in the fit. Third, the practical implementation of the DGLAP evolution from the initial scale $Q_0$ to the scale $Q$.

In fig.~\ref{fig:photonPDF} we show the photon PDF for the different sets at the scales $Q=(3,173, 5000)$ GeV. As it can be seen, these assumptions affect the dependence on $x$ and $Q$ for both the central value and the uncertainty band~\footnote{The red dotted curve labelled as {\sc\small APFEL\_NN23} will be explained after in the text.}. The main reasons for the differences can be traced to the different assumption for the photon PDF at the initial scale
and in differences in the QCD+QED evolution. In particular:
\begin{itemize}
    \item Consistently with the approach pursued for coloured partons, in {\sc\small NNPDF\-2.3\-QED} no functional form is specified for the photon PDF at the initial scale,  $\gamma(x,Q_0)$. The photon PDF is only constrained to be positive. In a first step, PDF replicas for all the partons are fit together from DIS-data only. Afterwards, they are further constrained by Drell-Yan data form the LHC Run-I at 7 TeV. At variance with DIS, neutral-current Drell-Yan is sensitive to the photon PDF already at LO, and it can put stronger constraints on $\gamma(x,Q_0)$. Because of the positivity requirement for the photon PDF, the replicas distribute in a very non-Gaussian way around the central value. The prescription suggested in order to determine a 68\% CL uncertainty band consists in the evaluation of the symmetric error including 68 of the 100 replicas around the central value. Since no model is assumed for the photon PDF and no data are present for large $x$, in this region uncertainties are very large and the central value alone can be misleading, leading even to an unphysical peak at very large $x$, which can be seen in fig.~\ref{fig:photonPDF}.
    \item The  {\sc\small CT14QED} and {\sc\small MRST\-2004\-QED} sets are based on a completely different assumption for $\gamma(x,Q)$ at the initial scale $Q=Q_0$.
            Since the amount of data for constraining the shape of the photon PDF is limited,  $\gamma(x,Q_0)$ is chosen to be described by an ansatz;  the photon PDF at $Q=Q_0$ is assumed to be completely determined by the valence quark distributions.  Specifically,  in {\sc\small CT14QED} and {\sc\small MRST\-2004\-QED}  the  photon parameterisation at the initial scale $Q_0$ GeV reads
            \begin{equation}
            f_{\gamma/p}(x,Q_0)\ =\ \frac{\alpha}{2\pi}\left(A_ue_u^2\tilde{P}_{\gamma q}\otimes u^0(x)+A_de_d^2\tilde{P}_{\gamma q}\otimes d^0(x)\right)\, . \label{eq:photonPDFintro}
            \end{equation}
            In eq.~\eqref{eq:photonPDFintro}
            $\tilde{P}_{\gamma q}\otimes f^0(x)$ corresponds to the convolution of the splitting function $\tilde{P}_{\gamma q}(x)$
            with the so-called ``primordial''
            quark distributions $f^0(x)$. 
            In the case of {\sc\small MRST\-2004\-QED}, $f^0(x)$ are valence-like model distributions, {\it i.e.}, they are not those fitted within the global {\sc\small MRST\-2004}~\cite{Martin:2004ir} analysis.            
            In the case of {\sc\small CT14QED}, $f^0(x)$ correspond to the initial up and down valence distributions 
            from {\sc\small CT14NLO}~\cite{Dulat:2015mca}.
            
            More importantly,  in {\sc\small CT14QED} $A_u$ is set equal to $A_d$ in order to obtain a dependence on a single parameter. 
              Conversely, in  {\sc\small MRST\-2004\-QED}, the coefficients $A_i$ are given by  $A_i=\ln\left(Q_0^2/m_i^2\right)$, where $m_i$ are the ``Current Mass'' (CM) of the quarks ($m_u=6$ MeV and $m_d=10$ MeV).             
          Thus,  the case of  {\sc\small CT14QED} is a generalisation of the original ansatz used in {\sc\small MRST\-2004\-QED}, where the photon PDF is simply determined by the CM for the up and the down quark. The {\sc\small CT14QED} set leaves freedom for the value of $m_i$ in the $A_i$ coefficients, which can be alternatively parameterised by the momentum fraction $p_\gamma^0$ carried by the photon at the initial scale.
The constraints on $p_\gamma^0$ and consequently on the photon PDF are then obtained by fitting the ZEUS 
    data~\cite{Chekanov:2009dq} for the production of isolated photons in Deep-Inelastic-Scattering (DIS), $ep\rightarrow e\gamma+X$, a process which is sensitive to the photon PDF already at LO.
    In this way, the constraint $p_0^\gamma\lesssim0.14\%$ is set at 90\% CL for the photon PDF, 
    at the initial scale of $Q_0=1.295$ GeV.
    For this reason, the uncertainty band in {\sc\small CT14QED} corresponds to the area between the $p_0^\gamma=0.14\%$ and $p_0^\gamma=0\%$ predictions; no central, or preferred 
    value is given in this approach.
    Since in {\sc\small MRST\-2004\-QED} the photon PDF is simply determined from the quark PDF assuming CM for the up and the down quark, no uncertainty band is provided. Moreover, for this set, no constraints from data enter in the determination of the photon PDF.
    \item All three PDF sets implement the DGLAP evolution at NLO QCD+LO QED accuracy. However, while the scale is evol\-ved simultaneously with the same value for the QCD and QED evolution in {\sc\small CT14QED} and {\sc\small MRST\-2004\-QED}, in  {\sc\small NN\-PDF\-2.3\-QED} the two scales are run independently. The effect from the different evolution can be seen by comparing the  plots in fig.~\ref{fig:photonPDF}, which display the same quantities at different factorisation scales. At $Q=3~\gev$ the central values of the three sets are different, although the {\sc\small CT14QED} and {\sc\small NNPDF\-2.3\-QED}  photons are well compatible within the uncertainty. At large  $Q$,  while   {\sc\small CT14QED} and {\sc\small MRST\-2004\-QED} converge to similar values, the prediction from {\sc\small NNPDF\-2.3\-QED} is different. This trend is particularly visible at small $x$ and for large scales, where the photon PDF mostly stems from the quark densities via the DGLAP evolution. However, it is worth to remind that a new correct implementation of the coupled QCD+QED DGLAP equations is  now also available in {\sc\small APFEL} \cite{Bertone:2013vaa} and has been described in \cite{Bertone:2015lqa} in the context of the determination of new sets of PDFs including also lepton members.~\footnote{During the writing of this article also a new set {\sc\small NNPDF3.0QED} has been released~\cite{Bertone:2016ume} and it includes the correct DGLAP running at (N)NLO QCD + LO QED accuracy.} In fig.~\ref{fig:photonPDF} we included also  a red dashed line labelled  {\sc\small APFEL\_NN23}, which corresponds to the central value for the set denoted as ``C2'' in \cite{Bertone:2015lqa}. This set   is the same of {\sc\small NNPDF\-2.3\-QED} at the initial scale $Q_0=1 ~\gev$, but it is different at larger scales due to the different treatment of the evolution.  Indeed, the  photon PDF in {\sc\small APFEL\_NN23} is very close to the one  of {\sc\small NNPDF\-2.3\-QED} only at large $x$, while for small $x$ and  large $Q$ it converges to the values of {\sc\small CT14QED} and {\sc\small MRST\-2004\-QED}. This proves that the discrepancy between the different sets at small $x$ and large $Q$ is completely driven by the different treatment of the DGLAP evolution. 
    \end{itemize}

In fig.~\ref{fig:aglumo} we show the  photon-gluon luminosity as function of the invariant mass $M$  for the three PDF sets considered at 8, 13 and 100 TeV, with the factorisation scale $Q$  set equal to $M$. In the first inset we show for every luminosity the ratio of their values at $Q=M$ and $Q=m_t$, while in the second(third) inset we show at $Q=M$($Q=m_t$)  the ratio of every luminosity over the corresponding central value with {\sc\small NNPDF\-2.3\-QED}. For  the entire range in $M$ considered at the three different proton-proton-collision energies, the {\sc\small CT14QED} and {\sc\small NNPDF\-2.3\-QED}  photon-gluon luminosities are barely compatible or not compatible at all. The slightly better compatibility with $Q=m_t$ is mainly due to the larger uncertainty band in {\sc\small NNPDF\-2.3\-QED}.

It is important to note that the behaviour of the photon PDF at small $x$ and large $Q$, the region where differences among the PDF sets are the largest (see fig.~\ref{fig:photonPDF}) and are determined by the different QCD+QED DGLAP evolution, is not of particular interest for our study. Indeed, as can be seen in fig.~\ref{fig:aglumo}, the {\sc\small APFEL\_NN23} $g \gamma $ luminosity is very close to the {\sc\small NNPDF\-2.3\-QED} one and inside its uncertainty band. The insensitivity of our study on the different treatment of QCD+QED DGLAP running is in principle expected for most of the predictions at the LHC, since large factorisation scales $Q$ are typically not correlated with small values of $x$. Moreover, at small values of $M$, where small values of $x$ can be probed, the PDFs of quarks and gluons are  much larger than the photon PDF, leading to a relative suppression of photon-initiated contributions.   
 
In order to determine at which values of $M$ the photon--gluon luminosity is actually expected to be  relevant in $t\bar{t}$ hadro\-production,  
in fig.~\ref{fig:luminosities} we show the photon-gluon, gluon-gluon and quark-antiquark luminosities  for {\sc\small NNPDF\-2.3\-QED} sets  at 8, 13 and 100 TeV. These plots will   also become useful  to understand the dependence of $\Sigma_{\rm EW}/\Sigma_{\rm LO~QCD}$ on the scale definition.~\footnote{In our calculation there are also contribution from quark-photon initial states, however, as it will be discussed in sect.~\ref{sec:res}, their contribution is small so we decided to exclude the corresponding luminosities from the discussion in this section for the sake of clarity.}

The solid lines refer to the luminosities with $Q=M$, with the uncertainty band shown only for the photon-gluon case. The same-colour dashed lines are the corresponding quantities 
with $Q=m_t$. The ratios between predictions with $Q=M$ and with $Q=m_t$ is shown in the first inset. The luminosities at $Q=M$ are in general of the same order or 
smaller than at $Q=m_t$, 
with the exception of the 100 TeV case for $M\lesssim 10~\tev$. In particular at 8 and 13 TeV, in the range of $M\lesssim 1~ \tev$ the luminosities of $gg$ and $q\bar{q}$ strongly reduce at $Q=M$ with respect to $Q=m_t$, while the photon-gluon luminosity is less sensitive to the value of $Q$. Since in the TeV range the negative contributions from Sudakov logarithms in the $gg$ and $q\bar q$ channels and the positive photon-gluon contribution tend to cancel each other, larger scales lead to larger values of $\Sigma_{\rm EW}/\Sigma_{\rm LO~QCD}$. This effect will be discussed both in sect.~\ref{sec:res} for predictions at   13 TeV and in sect.~\ref{sec:data}, where the $\Sigma_{\rm QCD+EW}/\Sigma_{\rm QCD}$ ratio will be compared to the experimental accuracy reached at the LHC measurements at 8 TeV.
In the second(third) insets we also show the ratios of the $gg$ and $q \bar{q}$ luminosities over the $g \gamma$ luminosity at $Q=M$($Q=m_t$) in order to better identify at which scales the $g \gamma$ contribution is expected to be relevant. From these insets, the hierarchy 
between $gg$ and $q \bar{q}$ channels can also be easily derived.

In conclusion, the different results for photon-induced processes and consequently for EW corrections obtained with the different PDF sets discussed in this article have to be attributed mainly to the different assumptions that underlie the determination of $\gamma(x,Q)$ at the initial scale $Q=Q_0$ and the choice of the factorisation scale, but  not to the different treatment of the DGLAP NLO QCD+LO QED evolution.

\section{Predictions at 13 and 100 TeV}
\label{sec:res}

\begin{figure*}
\centering
\includegraphics[width=0.45\textwidth]{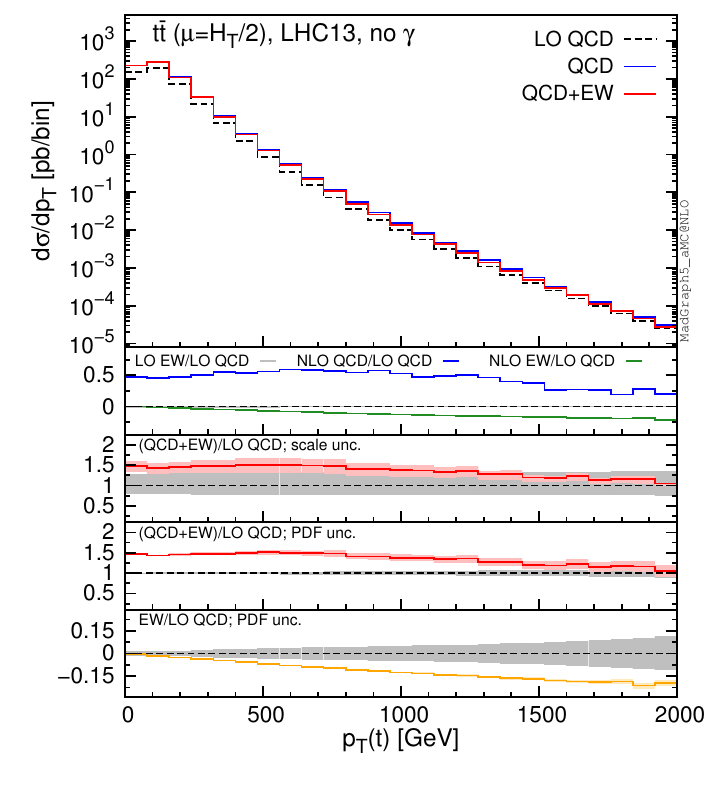}
\includegraphics[width=0.45\textwidth]{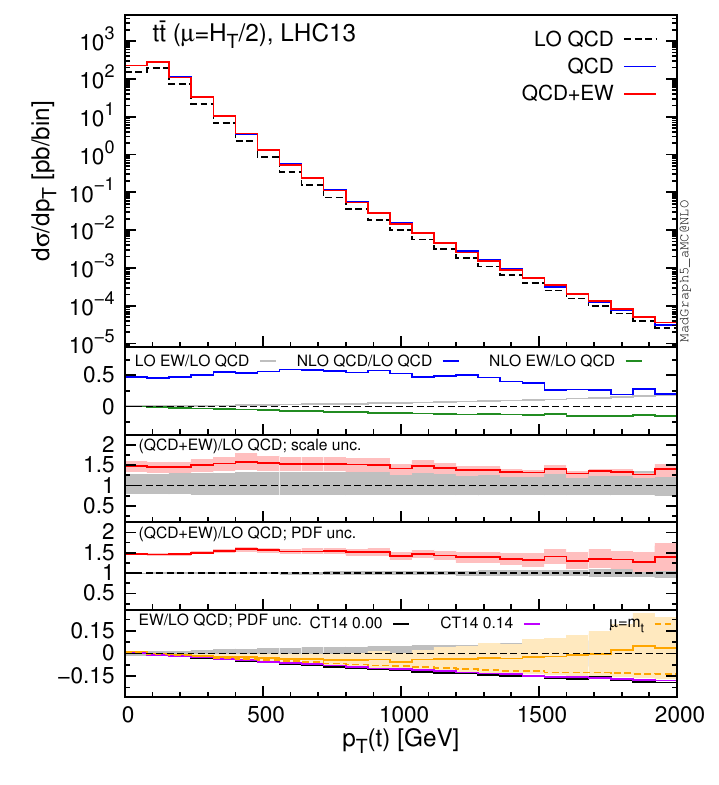}
\caption{Differential distributions for the $p_T(t)$ at 13 TeV. The format of the plots is described in detail in the text.}
\label{fig:LHC13_ptt}
\end{figure*}

\begin{figure*}
\centering
\includegraphics[width=0.45\textwidth]{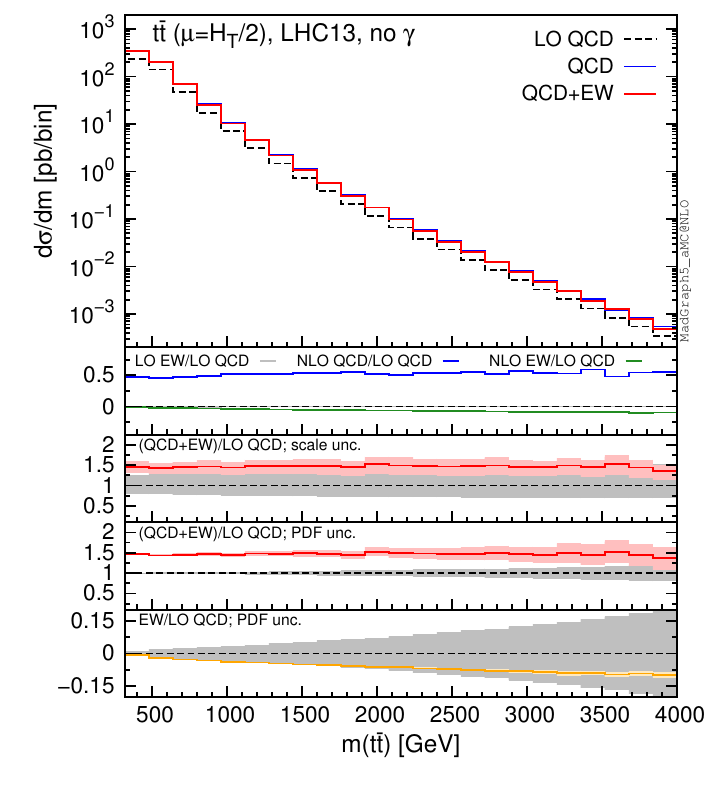}
\includegraphics[width=0.45\textwidth]{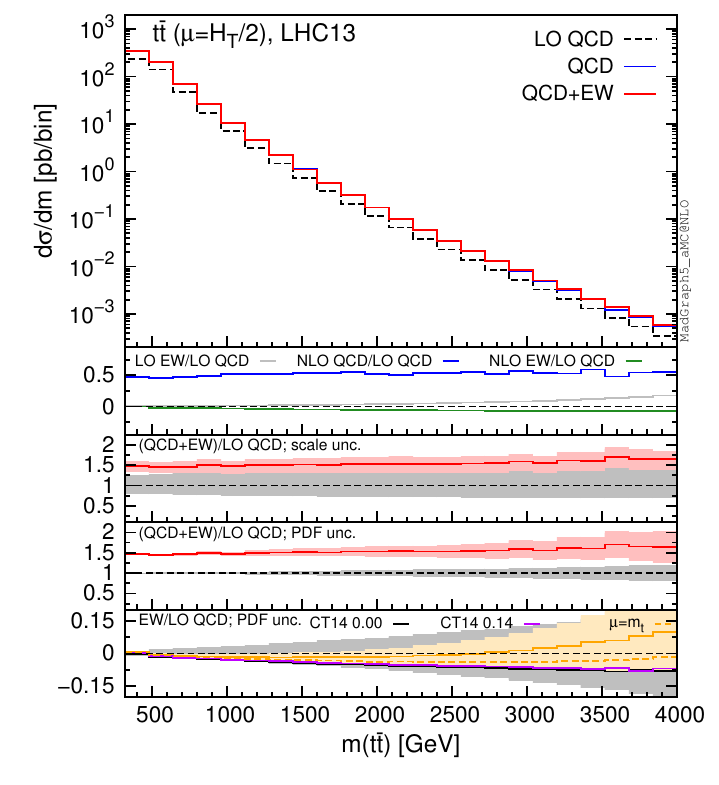}
\caption{Differential distributions for the $m(\ttt)$ at 13 TeV. The format of the plots is described in detail in the text.}
\label{fig:LHC13_mtt}
\end{figure*}

\begin{figure*}
\centering
\includegraphics[width=0.45\textwidth]{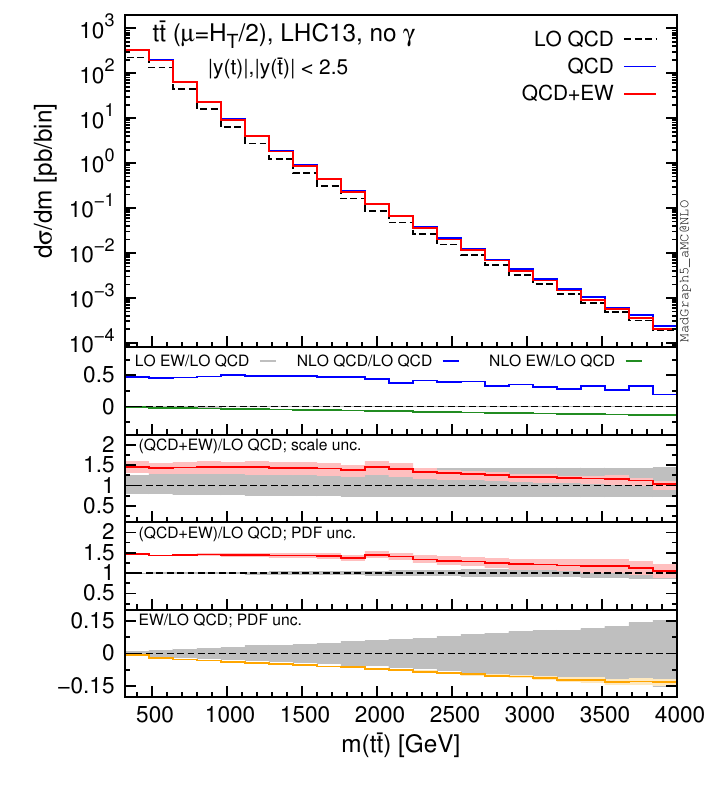}
\includegraphics[width=0.45\textwidth]{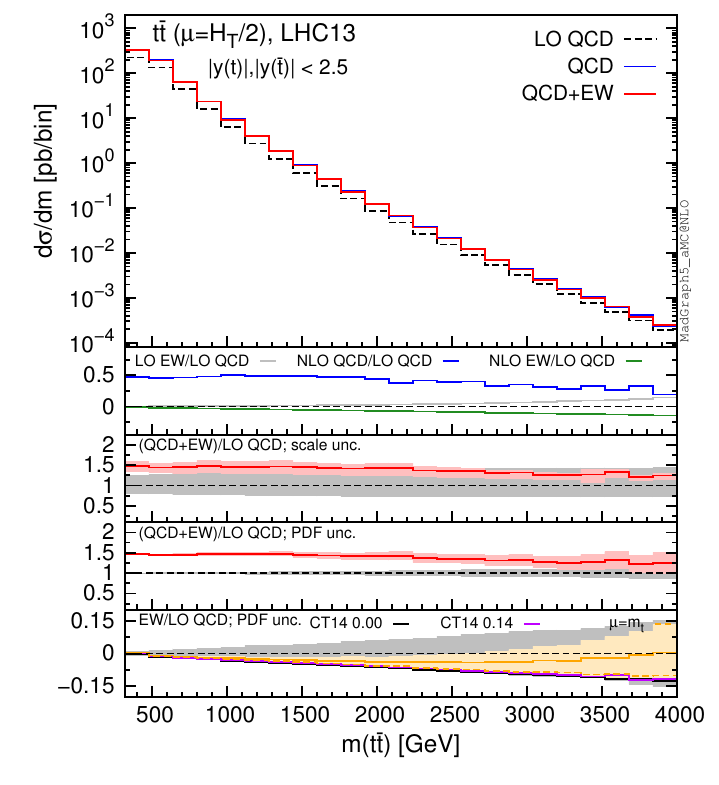}
\caption{Differential distributions for the $m(\ttt)$ at 13 TeV with $|y(t)|,|y(\bar t)|<2.5$ cuts applied. The format of the plots is described in detail in the text.}
\label{fig:LHC13_mtt_ycut}
\end{figure*}

\begin{figure*}
\centering
\includegraphics[width=0.45\textwidth]{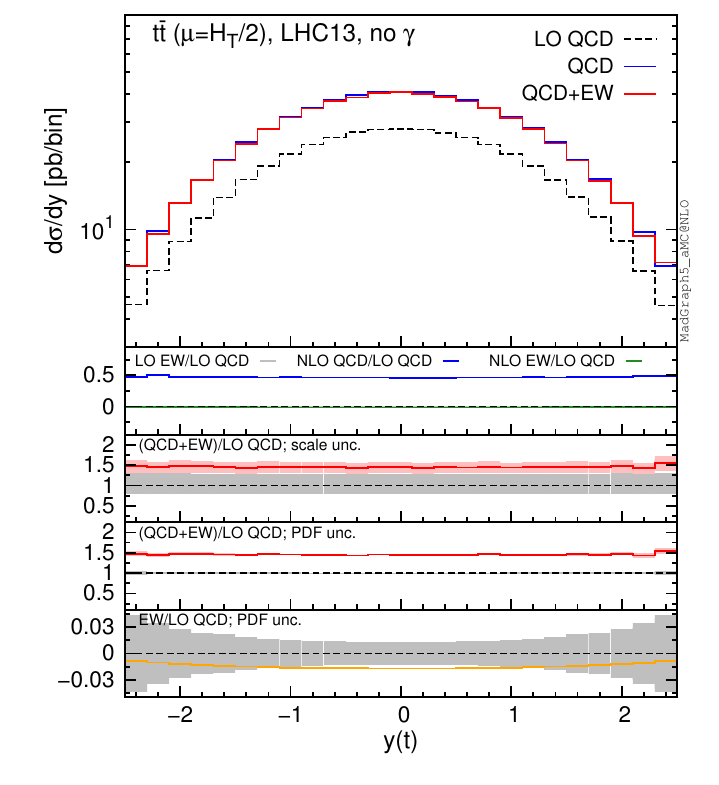}
\includegraphics[width=0.45\textwidth]{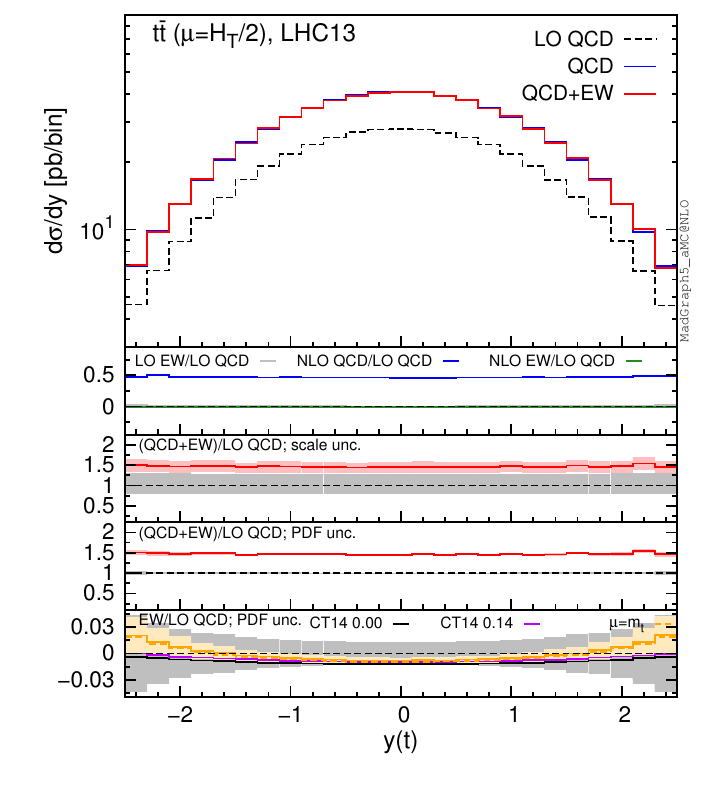}
\caption{Differential distributions for the $y(t)$ at 13 TeV. The format of the plots is described in detail in the text.}
\label{fig:LHC13_yt}
\end{figure*}

\begin{figure*}
\centering
\includegraphics[width=0.45\textwidth]{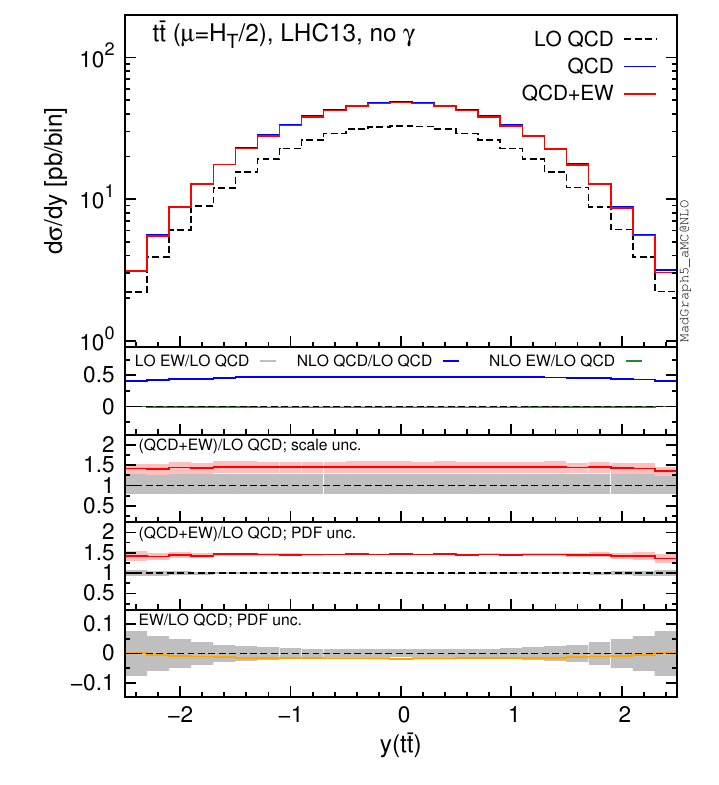}
\includegraphics[width=0.45\textwidth]{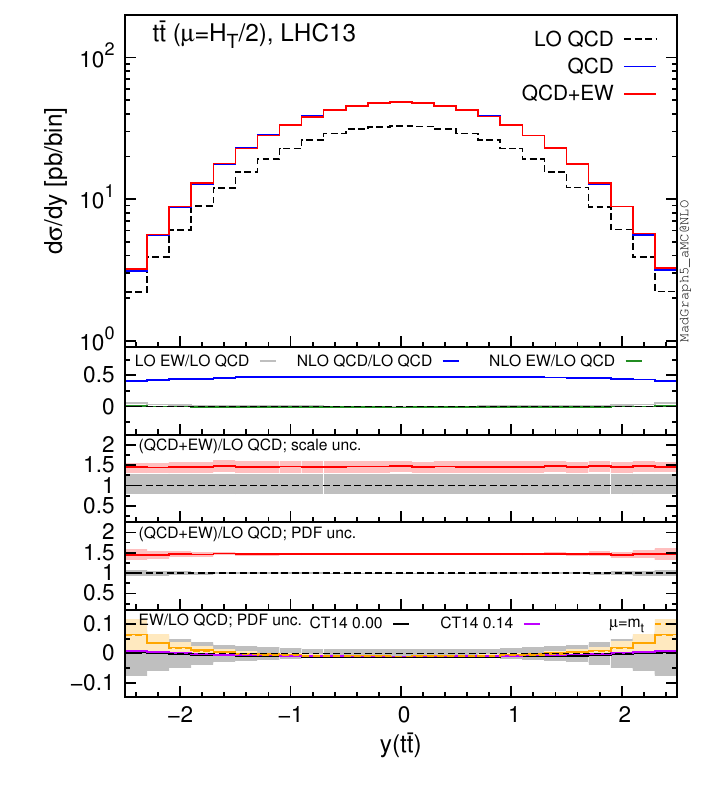}
\caption{Differential distributions for the $y(\ttt)$ at 13 TeV. The format of the plots is described in detail in the text.}
\label{fig:LHC13_yttx}
\end{figure*}

\begin{figure*}
\centering
\includegraphics[width=0.45\textwidth]{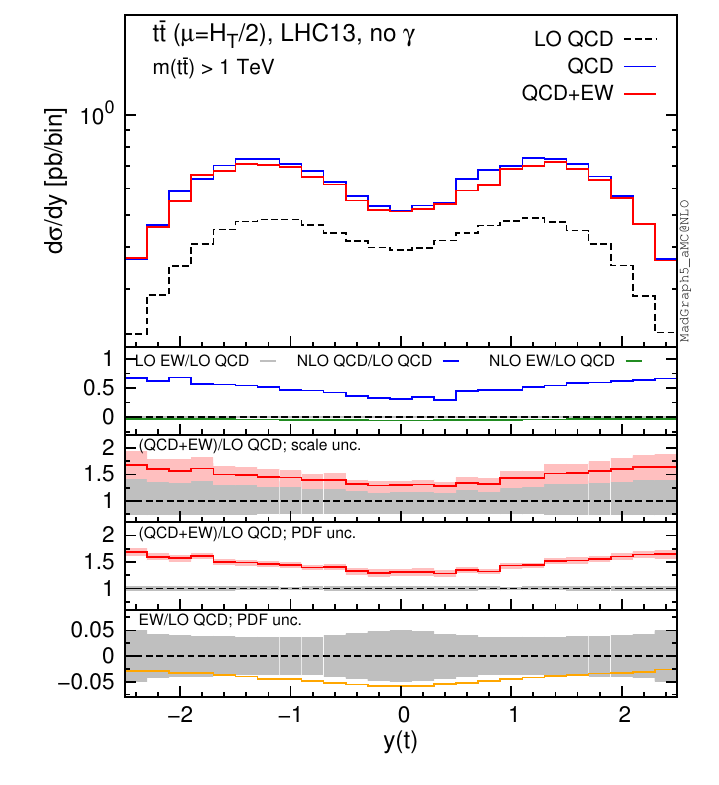}
\includegraphics[width=0.45\textwidth]{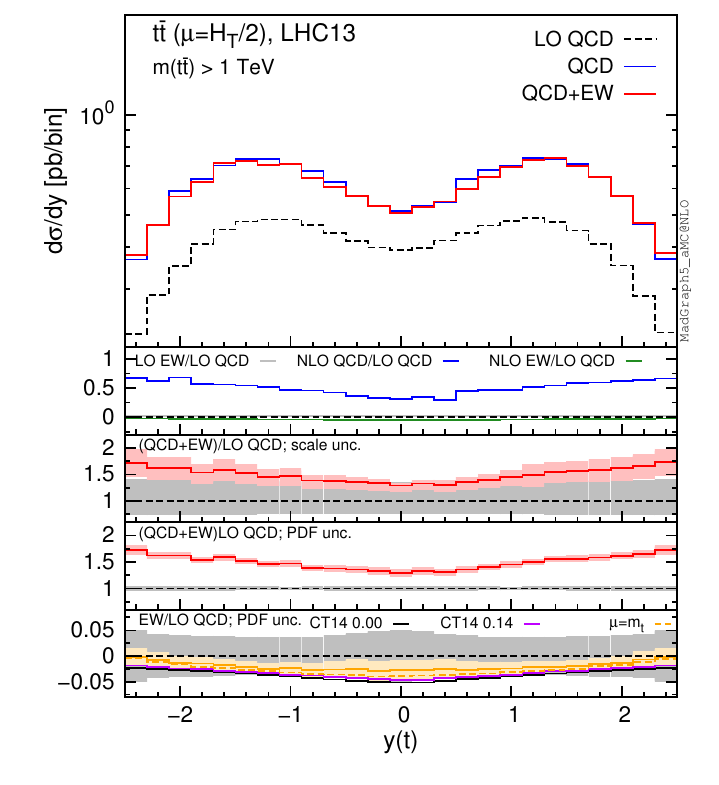}
\caption{Differential distributions for the $y(t)$ at 13 TeV with the cut $m(\ttt)>1~\tev$. The format of the plots is described in detail in the text.}
\label{fig:LHC13_yt_mttcut}
\end{figure*}

\begin{figure*}
\centering
\includegraphics[width=0.45\textwidth]{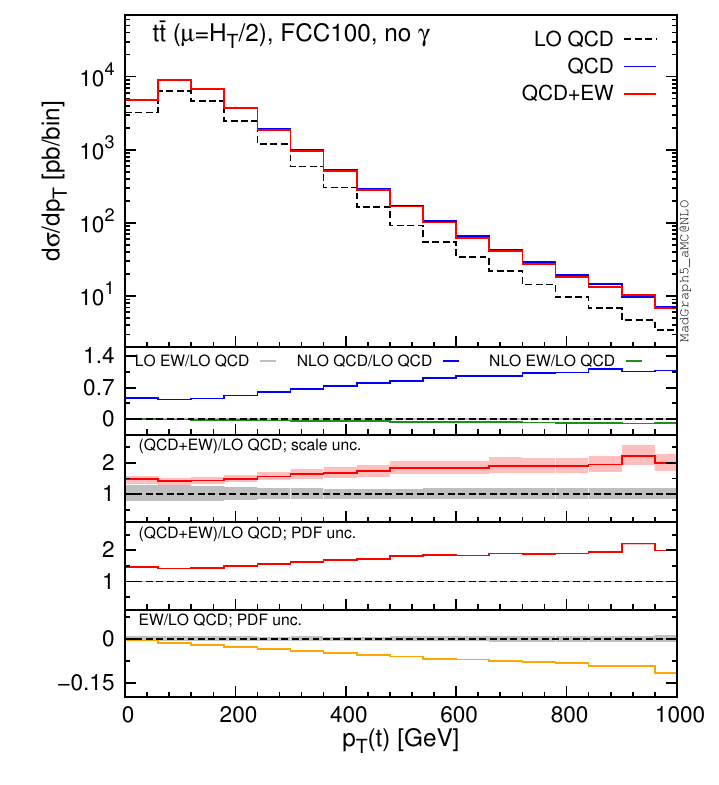}
\includegraphics[width=0.45\textwidth]{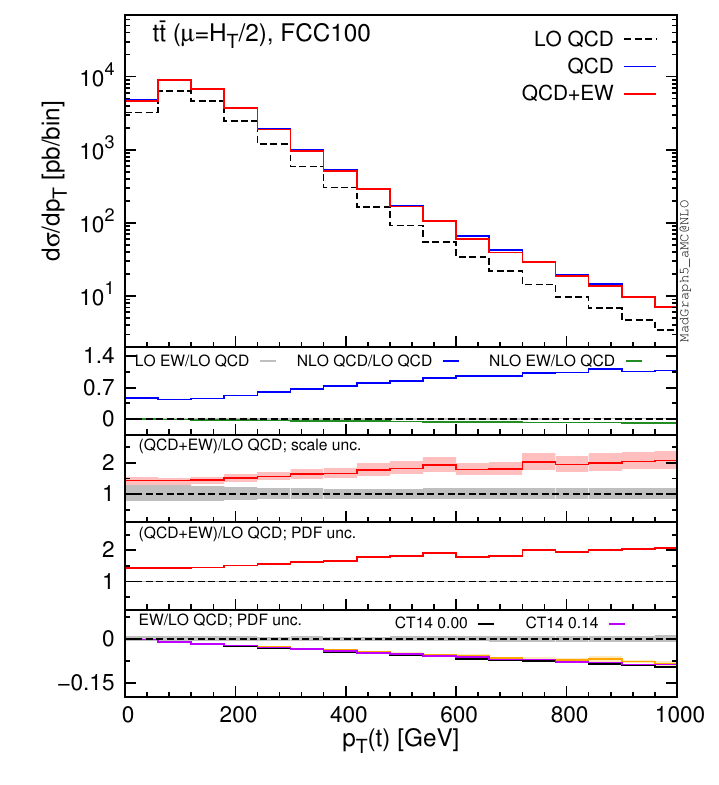}
\caption{Differential distributions for the $p_T(t)$ at 100 TeV. The format of the plots is described in detail in the text.}
\label{fig:LHC100_ptt}
\end{figure*}

\begin{figure*}
\centering
\includegraphics[width=0.45\textwidth]{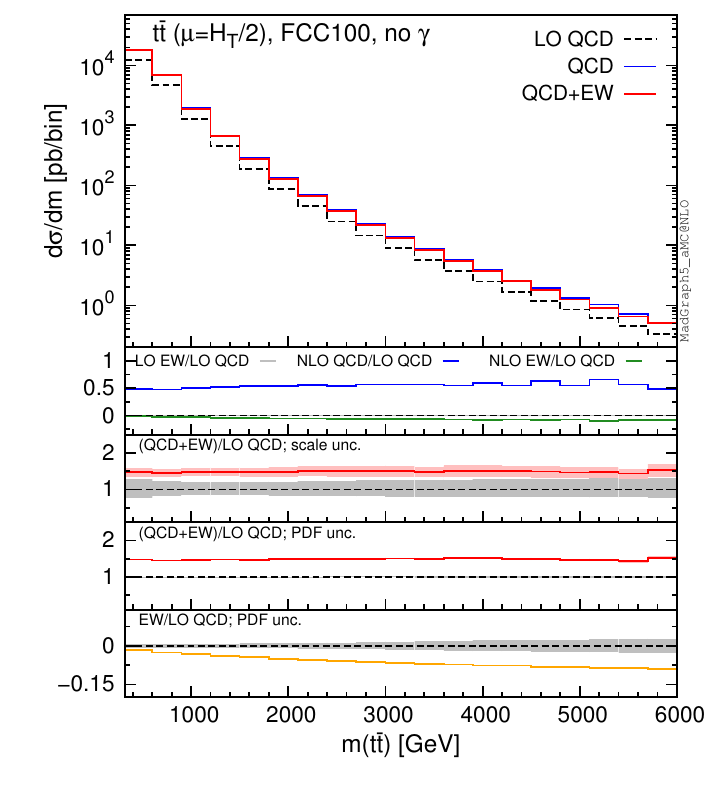}
\includegraphics[width=0.45\textwidth]{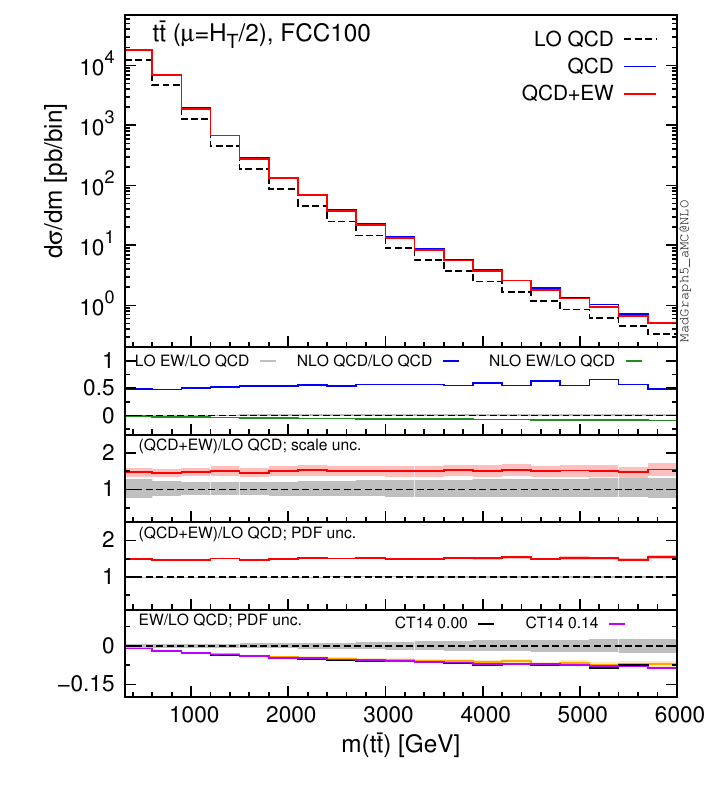}
\caption{Differential distributions for the $m(\ttt)$ at 100 TeV. The format of the plots is described in detail in the text.}
\label{fig:LHC100_mtt}
\end{figure*}

\begin{figure*}
\centering
\includegraphics[width=0.45\textwidth]{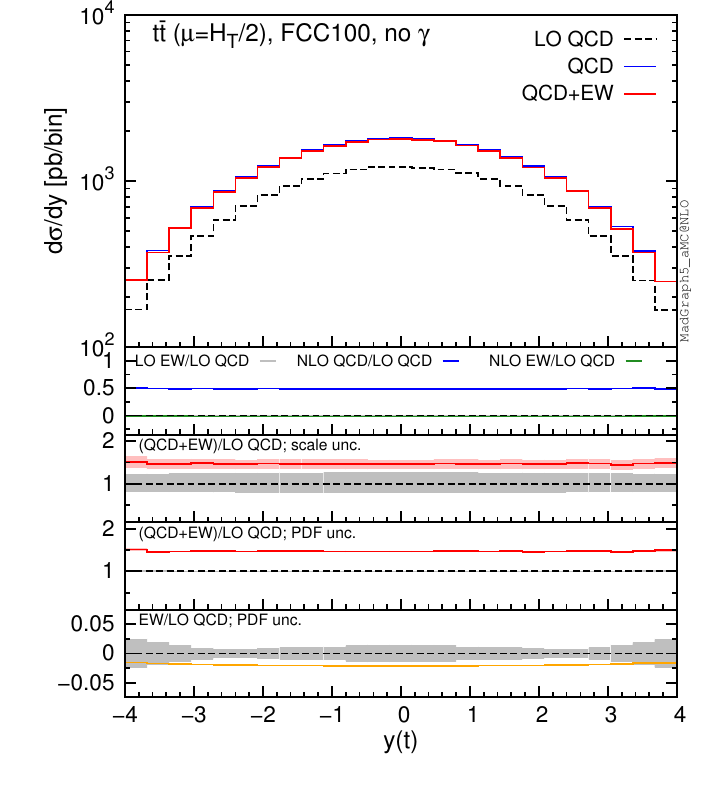}
\includegraphics[width=0.45\textwidth]{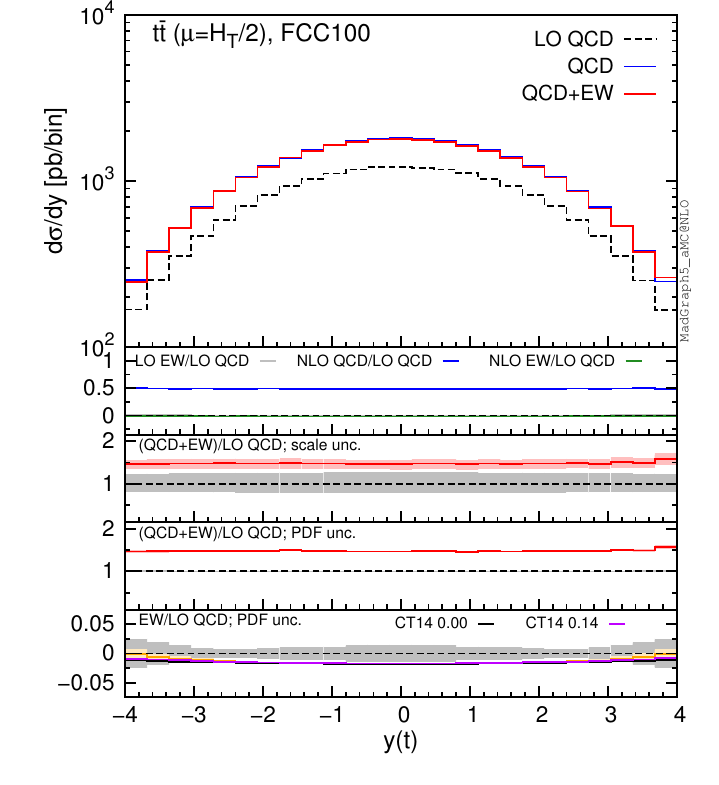}
\caption{Differential distributions for the $y(t)$ at 100 TeV. The format of the plots is described in detail in the text.}
\label{fig:LHC100_yt}
\end{figure*}

\begin{figure*}
\centering
\includegraphics[width=0.45\textwidth]{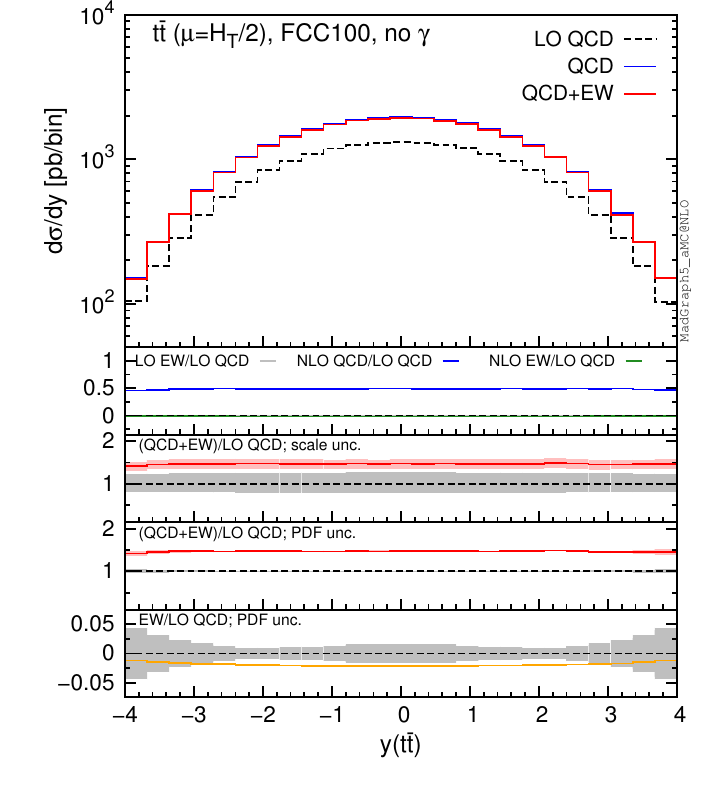}
\includegraphics[width=0.45\textwidth]{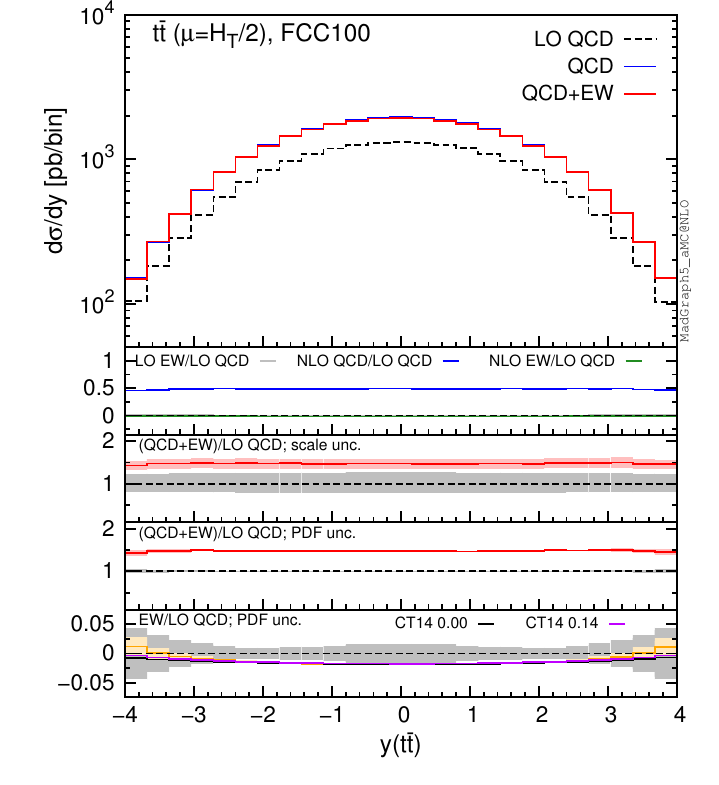}
\caption{Differential distributions for the $y(\ttt)$ at 100 TeV. The format of the plots is described in detail in the text.}
\label{fig:LHC100_yttx}
\end{figure*}

\begin{figure*}
\centering
\includegraphics[width=0.45\textwidth]{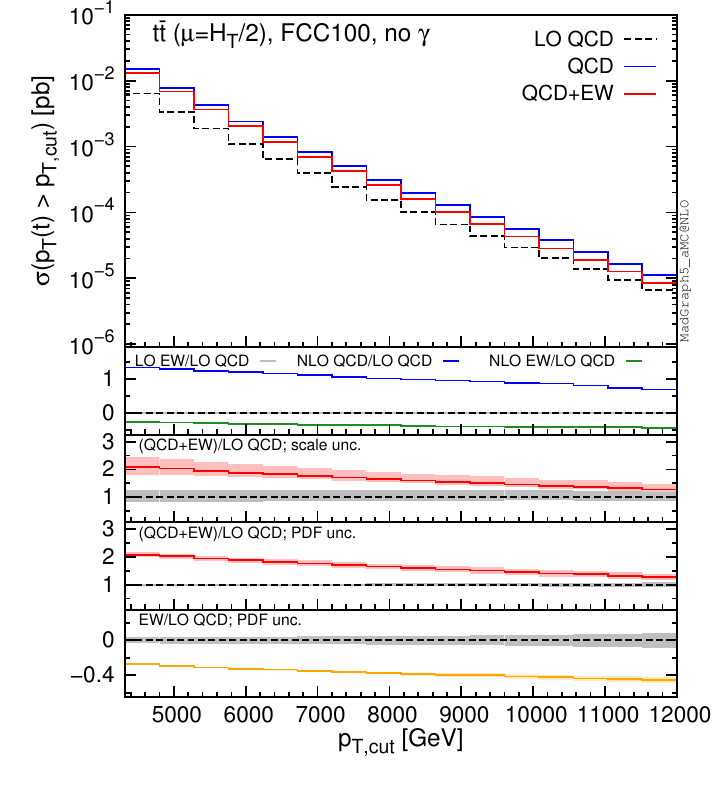}
\includegraphics[width=0.45\textwidth]{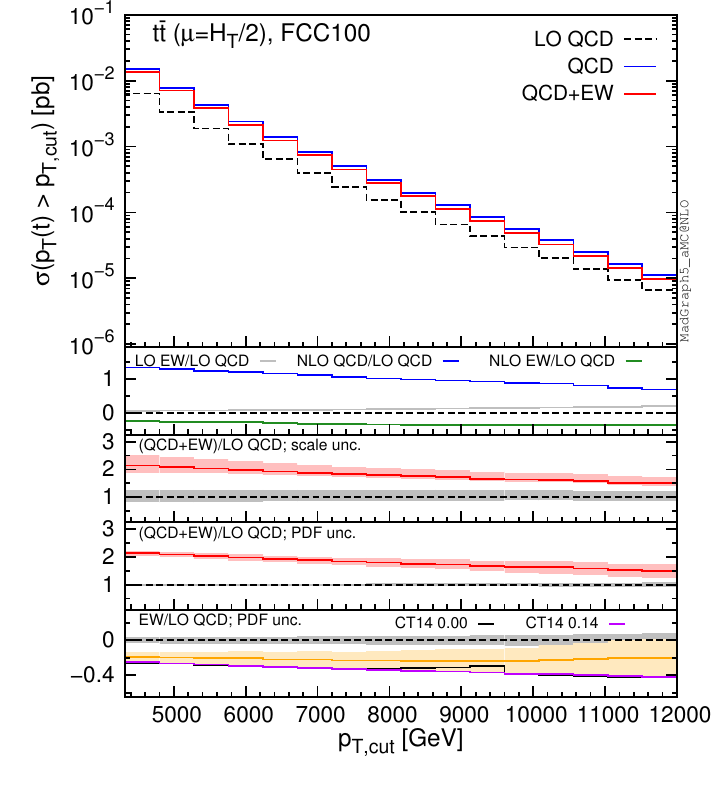}
\caption{Integrated distributions for the $p_T(t)$ at 100 TeV. The format of the plots is described in detail in the text.}
\label{fig:LHC100_pttcumu}
\end{figure*}

\begin{figure*}
\centering
\includegraphics[width=0.45\textwidth]{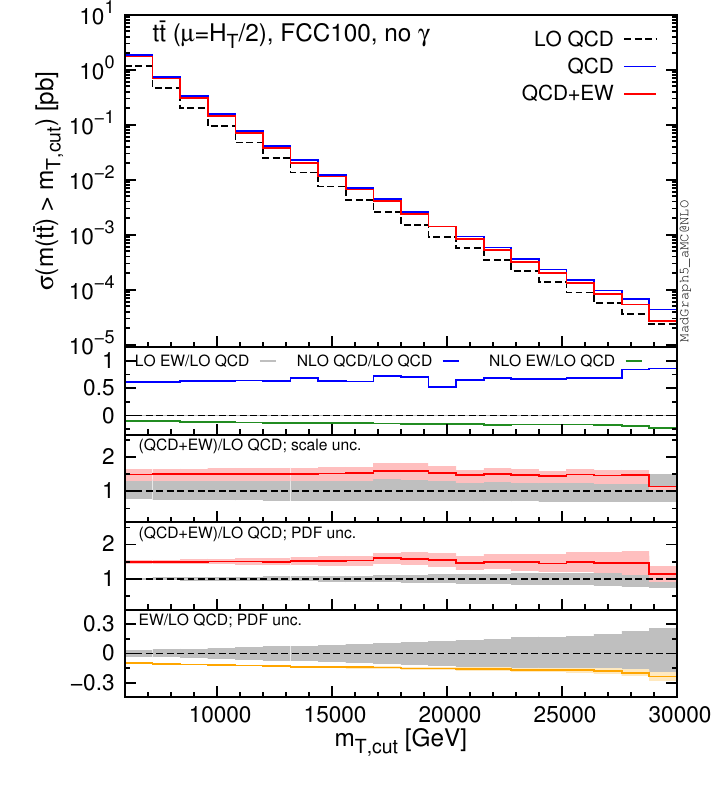}
\includegraphics[width=0.45\textwidth]{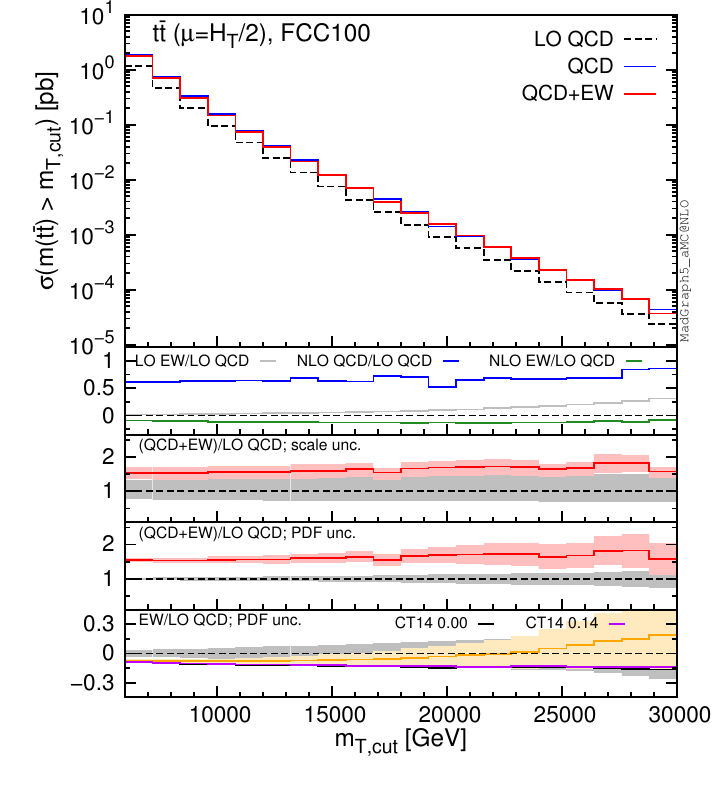}
\caption{Integrated distributions for the $m(\ttt)$ at 100 TeV. The format of the plots is described in detail in the text.}
\label{fig:LHC100_mttcumu}
\end{figure*}

In this section we discuss the impact of the EW corrections and the photon PDF on several distributions  at 13 and 100 TeV. In particular, we focus on the top-pair invariant mass $m(\ttt)$, the transverse momentum of the top  quark $ p_T(t)$, and the rapidities of the top  quark  $y(t)$ and $\ttt$ pair $y(\ttt)$. Predictions for the LHC at 13 TeV are shown in figs.~\ref{fig:LHC13_ptt}-\ref{fig:LHC13_yt_mttcut}, while those for a Future Circular Collider (FCC) at 100 TeV are shown in figs.~\ref{fig:LHC100_ptt}-\ref{fig:LHC100_mttcumu}.  Unless differently specified,  results are obtained with no cut imposed on the final-state particles.

In each figure 
we show two plots for the same observable, displaying in the left plot, denoted as ``no $\gamma$'', predictions  with  the photon PDF artificially set equal to zero.
 The reason behind this choice is manifold. First of all, the comparison of left and right plots allows to gauge the impact of the photon PDF on both the central value and PDF uncertainties of the electroweak contributions. Furthermore, in the plots on the left it is possible to observe the impact of the Sudakov logarithms, which can be hidden by the compensation given by the photon-induced processes included in the plots on the right. Last but not least, whereas the EW corrections in the right plots strongly depend on the PDF set used and may possibly change with  future improved determination of the photon PDF,  the plots on the left display the subset of the EW corrections that is expected to be stable under future PDF determination.    

In each plot the main  panel includes the distributions for $\Sigma_{\rm LO~QCD}$, $\Sigma_{\rm QCD}$ and $\Sigma_{\rm QCD+EW}$ predictions as defined in equations \eqref{blobbegin}-\eqref{blobend}. The four insets below the main panel display ratios of several quantities over the $\Sigma_{\rm LO~QCD}$ central value. In the first inset we plot this ratio for the central values of $\Sigma_{\rm LO~EW}$, $\Sigma_{\rm NLO~QCD}$ and $\Sigma_{\rm NLO~EW}$, {\it i.e.}, the other perturbative orders which we consider. The second and third inset respectively present the $\Sigma_{\rm QCD+EW}/\Sigma_{\rm LO~QCD}$  ratio including the scale and PDF uncertainties for the numerator. In the last inset we present the $\Sigma_{\rm EW}/\Sigma_{\rm LO~QCD}$ ratio with the PDF uncertainties for the numerator. All the results in the plots are obtained with the {\sc\small NNPDF\-2.3\-QED} PDF set, with the PDF uncertainty computed at the 68\% CL. However, plots on the right, including
effects due to the photon PDF, also show (in the last inset) the ratio $\Sigma_{\rm EW}/\Sigma_{\rm LO~QCD}$ computed with the {\sc\small CT14QED} PDF set.~\footnote{Only for this ratio, also the $\Sigma_{\rm LO~QCD}$ quantity has been evaluated with the {\sc\small CT14QED} PDF set.} For this particular set two predictions are shown, which correspond to a momentum fraction carried by the photon at the initial scale of 0\% and 0.14\%; as explained in sect.~\ref{sec:PDF} this interval corresponds to the uncertainty band of  {\sc\small CT14QED}.

The first general feature that can be noticed in all the distributions for the LHC at 13 TeV is the different  behaviour of the EW corrections with {\sc\small NNPDF\-2.3\-QED} and {\sc\small CT14QED} in the plots on the right. The effects induced by the photon PDF with {\sc\small CT14QED} is much smaller w.r.t. the case of   {\sc\small NNPDF\-2.3\-QED}.
Indeed, by comparing the last inset in equivalent plots with (right) and without (left) photon-induced contributions, it is easy to see that effects due to the {\sc\small CT14QED} photon PDF are essentially invisible, regardless of the momentum fraction (in the 0\%-0.14\% range). Furthermore, by comparing the first inset in the two plots of each figure, it is possible to notice how the dominant contribution from photon-induced processes originates from the $\Sigma_{\rm LO~EW}$ contribution of the $g \gamma $ channel. As anticipated in sect.~\ref{sec:PDF}, the $q \gamma $ contribution is negligible~\footnote{This statement could be done  only {\it a posteriori}, after having performed the calculation and cannot be generalised to other processes. See for instance ref.~\cite{Baglio:2013toa}.  } and will not be further discussed in the following. Moreover, when we will refer to the effect of photon PDF, unless differently specified, we will always understand the case of {\sc\small NNPDF\-2.3\-QED} PDF set in order to be conservative.

As can be seen in figs.~\ref{fig:LHC13_ptt} and \ref{fig:LHC13_mtt}, at 13 TeV the photon-initiated contributions can  compensate the Sudakov suppression in the top-pair invariant mass and top-quark transverse momentum distributions. At very large scales  ($p_T(t) > ~ 1.7~\tev$ or $m(\ttt) > ~ 3~\tev$)  their size is larger than the effect due to Sudakov logarithms. In all the figures  at 13 TeV, in the last inset of the plots on the right, we also included  the central value of the ratio $\Sigma_{\rm EW}/\Sigma_{\rm LO~QCD}$ with both quantities evaluated at the scale $\mu=m_t$. This quantity indicates that the cancellation between  Sudakov logarithms strongly depends also on the scale choice, thus  the scale dependence of $\Sigma_{\rm EW}$ should also be taken into account as a source of uncertainties. The origin of this dependence will be discussed in detail in sect.~\ref{sec:data}, in the context of the measurements at 8 TeV. 

In the last inset, the PDF uncertainties of the $\Sigma_{\rm EW}$ term alone are directly compared  with the $\Sigma_{\rm LO~QCD}$ PDF uncertainties, the grey band.
In the tail of the  $p_T(t)$ and $m(\ttt)$ distributions on the right, PDF uncertainties for separately $\Sigma_{\rm EW}$ and $\Sigma_{\rm LO~QCD}$ are of the same size. This is consistent with the difference between the PDF uncertainty   for the total prediction $\Sigma_{\rm QCD+EW}$ with and without effects form photon in the initial state, which is shown  in the third inset and, in the plots on the right, is much larger w.r.t the case of $\Sigma_{\rm LO~QCD}$. Just for reference, we decided to show in the second inset the ratio $\Sigma_{\rm QCD+EW}/\Sigma_{\rm LO~QCD}$  with scale uncertainties, which are reduced w.r.t. the $\Sigma_{\rm LO~QCD}$ case, the grey band. The effects due to the EW corrections and the impact of the photon PDF is in general smaller than the scale uncertainty of $\Sigma_{\rm QCD+EW}$. However, one should consider that the scale uncertainty is further reduced to very few percents by NNLO corrections, which have already been calculated~\cite{Czakon:2013goa, Czakon:2015owf}. Moreover,  when normalised distributions ($d\sigma/\sigma$) are considered, even at NLO accuracy scale uncertainties are strongly reduced.  In sect.~\ref{sec:data} we will comment both effects for specific distributions measured at the LHC at 8 TeV by ATLAS and CMS, where we will also investigate the impact of EW corrections and of the photon PDF. 

While $p_T(t)$ distributions do not change by requiring $|y(t)|$, $|y(\bar t)| < 2.5$, {\it i.e.} top and anti-top produced in the central region of the detector, these cuts have an effect on predictions for high $m(t\bar t)$. This is shown in  fig.~\ref{fig:LHC13_mtt_ycut}, which is equivalent to fig.~\ref{fig:LHC13_mtt}, but with $|y(t)|,|y(\bar t)| < 2.5$ cuts. At large invariant masses, tops are preferably produced in the forward or backward region, due to the $t$- and $u$-channel diagrams in the $gg$-channel, which is much less suppressed than the $q\bar q$ channels featuring only $s$-channel diagrams at LO QCD. Rapidity cuts suppress the $gg$ contribution, as well as $g\gamma$ contributions, but also enhance the typical value of the partonic Mandelstam variables $\hat t$ and $\hat u$. Consequently, with those cuts, we observe larger values for the Sudakov logarithms (plot on the left) and a similar behaviour for the photon-induced contributions. Moreover, we can notice that, as in the $p_T(t)$ distributions in fig.~\ref{fig:LHC13_ptt}, also in the tails of the plots in fig.~\ref{fig:LHC13_mtt_ycut} the ratio $\Sigma_{\rm NLO~QCD}/\Sigma_{\rm LO~QCD}$ in the first insets decrease, at variance with  fig.~\ref{fig:LHC13_mtt}, where cuts are not applied. This trend is correlated with the fraction of the cross section originating from the $g g$ initial state, which is decreasing in the tail of $p_T(t)$ distribution and, only with rapidity cuts applied, of the $m(t \bar t)$ distribution.

In figs.~\ref{fig:LHC13_yt} and \ref{fig:LHC13_yttx} we show predictions for $y(t)$ and $y(\ttt)$ distributions. EW corrections are quite flat in the plots on the left, while in the plots on the right they are enhanced by the $g \gamma $ initial state in the peripheral region, especially for the $y(\ttt)$ distribution.  In this region photon-initiated contributions and their uncertainties become relevant since the photon PDF is sampled at rather large $x$, where the suppression w.r.t  the quark and gluon PDFs is reduced. At variance with the tail of $p_T(t)$ or $m(\ttt)$, where the $g \gamma $ contribution is also sizeable, the cross section in the peripheral region is not largely suppressed w.r.t to the total cross section. For these reasons, large-rapidity bins can be exploited to set additional constraints on the photon PDF in a theoretical framework such as the one adopted in {\sc\small NNPDF\-2.3\-QED}, while with an assumption {\it \`a la} {\sc\small CT14QED} the sensitivity on the photon PDF remains very small. Given the large cross section in $\ttt$ production at the LHC 13 TeV, in fig.~\ref{fig:LHC13_yt_mttcut} we also show the same plots of fig.~\ref{fig:LHC13_yt} with a $m(\ttt)>1~\tev$ cut applied. EW corrections in the left plot are larger in the central region, where, due to large values of the $\hat t$ and $\hat u$ Mandelstam variables, the effect of Sudakov logarithms is enhanced. Again, this effect can be compensated by photon-induced processes as shown in the plot on the right. As already observed for all the previous distributions, only in the plots on the left the PDF uncertainties related to $\Sigma_{\rm EW}$ are negligible, while they are sizeable in the plots on the right. Also in this case the trend of the ratio $\Sigma_{\rm NLO~QCD}/\Sigma_{\rm LO~QCD}$ displayed in the first inset is correlated with the fraction of the cross section originating from the $q \bar q$ initial state, as discussed for  figs.~\ref{fig:LHC13_ptt} and~\ref{fig:LHC13_mtt_ycut}.

It is worth to notice that while plots in fig.~\ref{fig:LHC13_ptt} would be identical in the case of $p_T(\bar t)$ distributions, the transverse momentum of the antitop, the $y(\bar t)$ distribution is different from $y( t)$ when  $\Sigma_{\rm QCD+EW}$ terms are included, because of the charge asymmetry \cite{Kuhn:1998kw,Hollik:2011ps,Kuhn:2011ri,Bernreuther:2012sx}.
However, we observed the same qualitative behaviour for the photon PDF and the EW corrections in $y(t)$ and $y(\bar t)$ distributions. Moreover, the $g \gamma $ channel cannot give a contribution to the numerator of the charge asymmetry (see also ref.~\cite{Pagani:2012kj}), and all the other contributions to the asymmetry have already been investigated in ref.~\cite{Bernreuther:2012sx}.

The distributions that have been discussed for the LHC at 13 TeV are also presented for a FCC at 100 TeV, with larger ranges for the abscissae.~\footnote{We provided a few representative results also  in \cite{Mangano:2016jyj}.} By comparing plots in figs.~\ref{fig:LHC100_ptt}-\ref{fig:LHC100_yttx} with their corresponding ones at 13 TeV, it can be noticed that the impact of the photon PDF is strongly reduced at 100 TeV. In each figure, the plot on the right (with photons in the initial state) does not exhibit any qualitatively different behaviour w.r.t. the plot on the left. The smaller impact of the photon-induced contributions at 100 TeV w.r.t the 13 TeV case is due to the different range of $x$ spanned in the PDFs; keeping the hardness of the process fixed, a larger energy of the hadronic collisions corresponds to probing smaller values of  $x$, where parton luminosities
involving photons are suppressed with respect to those involving QCD partons, as shown in fig.~\ref{fig:luminosities}. For the same reason, the impact of the photon PDF at the LHC at 8 TeV is even larger than at 13 TeV, as it will be discussed in the next session. Moreover, at 100 TeV,  for a given value of $p_T(t)$ or $m(\ttt)$, EW corrections are slightly smaller than at 13 TeV also in the plots on the left, {\it i.e.}, without considering photon-induced contributions. 

In order to find a large contribution form the $g \gamma $ initial state at a FCC, it is necessary to probe very large scales. In  figs.~\ref{fig:LHC100_pttcumu} and \ref{fig:LHC100_mttcumu} we respectively plot, using the same layout of the previous plots, the cumulative distributions  $\sigma(p_T(t)>p_{T,{\rm cut} })$ and $\sigma(m(\ttt)>m_{{\rm cut} })$ with $p_{T,{\rm cut}}$   up to 12 TeV and  $m_{{\rm cut} }$ up to 30 TeV. First of all, we remind that at a FCC  it would be possible with $10~{\rm ab}^{-1}$ integrated luminosity to probe such small cross sections and large hard-scales. Thus, the following discussion is not  purely academic.  Sudakov logarithms can induce up to $\sim -40\%$ corrections for $\sigma(p_T(t)> 12~\tev)$ and  $\sim -30\%$  for $\sigma(m(\ttt)> 30~\tev)$, rendering EW corrections not only important for precision physics, but  essential for obtaining a sensible result. For precision physics, Sudakov logarithms have to be considered also beyond the NLO EW accuracy and possibly resummed. The central role of EW corrections at a FCC at 100 TeV is not peculiar for the $\ttt$ final state;  it is a general feature for all the production processes (see also the EW section in \cite{Mangano:2016jyj}). At such large scales, also the $g \gamma $ luminosity is not negligible, and indeed the size and the PDF uncertainty of the EW corrections is very different in the left and right plots. Once again, the prediction obtained with {\sc\small CT14QED} is similar to the case where the photon PDF has been set to zero. As in the case of 13 TeV, we observed similar changes in the tail of the $m(\ttt)$ distribution by applying the $|y(t)|,|y(\bar t)| < 4$ cuts, {\it i.e.}, mimicking the expected coverage of the future detector. Sudakov logarithms are enhanced, photon contributions are not visibly changed, and the $\Sigma_{\rm NLO~QCD}/\Sigma_{\rm LO~QCD}$ ratio in the first inset decreases in the tail as observed in fig.~ \ref{fig:LHC100_pttcumu}.

\section{Impact on ATLAS and CMS measurements at 8 TeV}
\label{sec:data}
   
In this section we discuss the relevance of the photon PDF and EW corrections for experimental measurements at the LHC.  Specifically, we will consider their impact on the analyses already published both by ATLAS and CMS for the $\ttt$ differential distributions at 8 TeV. 

We want to stress that our aim is not to perform a direct comparison of the SM ``best prediction'' with experimental data. Indeed, at differential level, the best accuracy reached in perturbative QCD is NNLO or NLO+NNLL. Terms beyond NLO in QCD are not taken into account in this analysis, but they are crucial both for a correct determination of the central value and for reducing the scale uncertainty. Here, we want  to show that a reliable comparison between SM predictions and experimental data in $t \bar{t}$ production cannot be based only on purely QCD corrections. Indeed, the contribution of EW corrections, especially the uncertainty due to the photon PDF and the scale choice, cannot be ignored and has to be taken into account. In particular cases, such as the normalised distributions discussed in the following, they can yield the largest effect, playing a primary role in the determination of precise  SM predictions.   

After the discussion and the results presented in the previous section, it is clear that  EW corrections at 8 TeV are expected to show a larger dependence on the photon PDF w.r.t. the 13 TeV case. On the other hand, smaller cross sections and integrated luminosities may in practice not lead to enough events to probe this effect. In order to evaluate their real impact at the experimental level, we take in to account in our analysis both the theory uncertainties  (scale  and  PDF)  and the errors from the experimental measurements. Moreover, this study serves also as a motivation to perform this kind of analysis for future 13 TeV data, where larger cross sections and luminosities should considerably decrease the experimental errors.

In the following, we consider CMS data  from ref.~\cite{Khachatryan:2015oqa}, which are based on the dilepton and lepton+jets events collected with an integrated luminosity of $19.7\ifb$. In particular, we focus on the distributions of the transverse momentum and rapidity of the top quark,~\footnote{In the plots of this section, the $p_T(t)$ distribution is actually the average of the distribution of the transverse momentum of the top quark and antiquark, consistently with what has been done in the experimental analyses. Similarly, the $y(t)$ distribution is the average of the distribution of the rapidity of the top quark and antiquark. However, differences with real $p(t)$ and $y(t)$ distributions are in practice invisible.} and of the rapidity of the $\ttt$ pair.
Moreover, we consider ATLAS data based on lepton+jets events collected with an integrated luminosity of $20.3\ifb$. Distributions for the top quark transverse momentum in the boosted regime are taken from ref.~\cite{Aad:2015hna}, while those for the invariant mass of the $\ttt$ pair, the absolute value of the rapidity of the top and of the $\ttt$ pair are taken from ref.~\cite{Aad:2015mbv}.
In the ATLAS analyses, part of the electroweak corrections (the purely weak  contribution calculated in \cite{Kuhn:2006vh}) has been taken into account via an updated version of {\sc\small Hathor} \cite{Aliev:2010zk}. Thus, although the Sudakov enhancement has been taken into account, the QED corrections and especially the dependence on the photon PDF, which is a crucial aspect of our discussion, have not been considered yet.

\begin{figure*}
\centering
\includegraphics[width=0.45\textwidth]{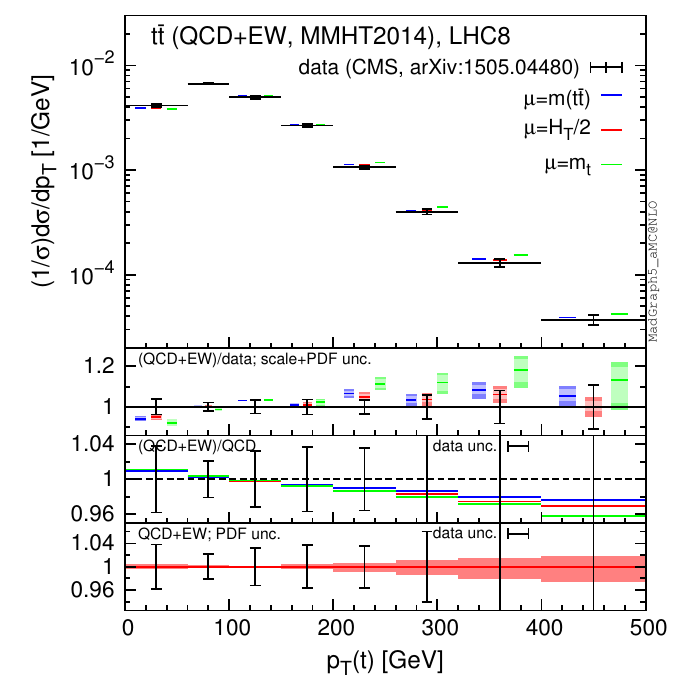}
\includegraphics[width=0.45\textwidth]{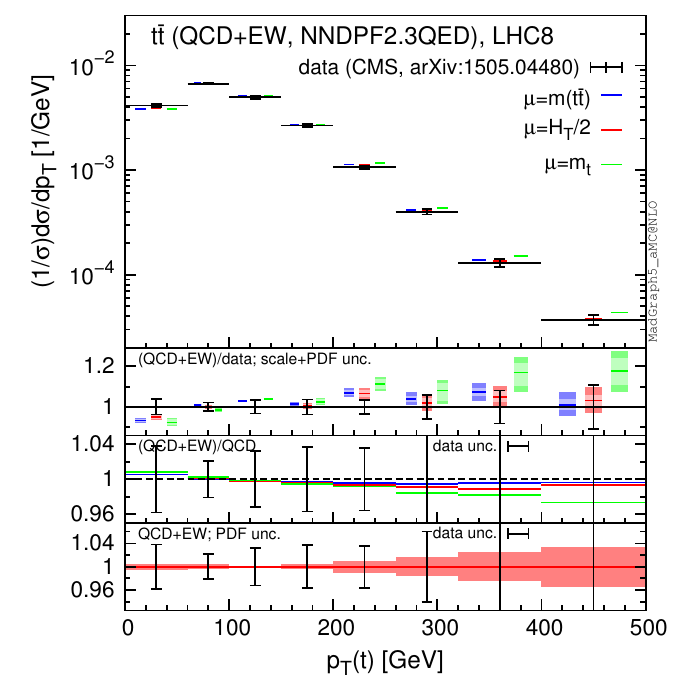}\\
\includegraphics[width=0.45\textwidth]{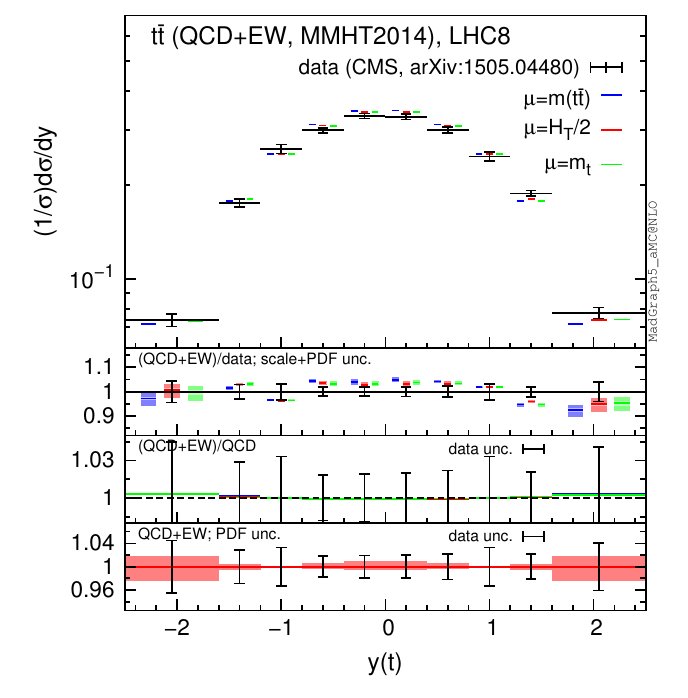}
\includegraphics[width=0.45\textwidth]{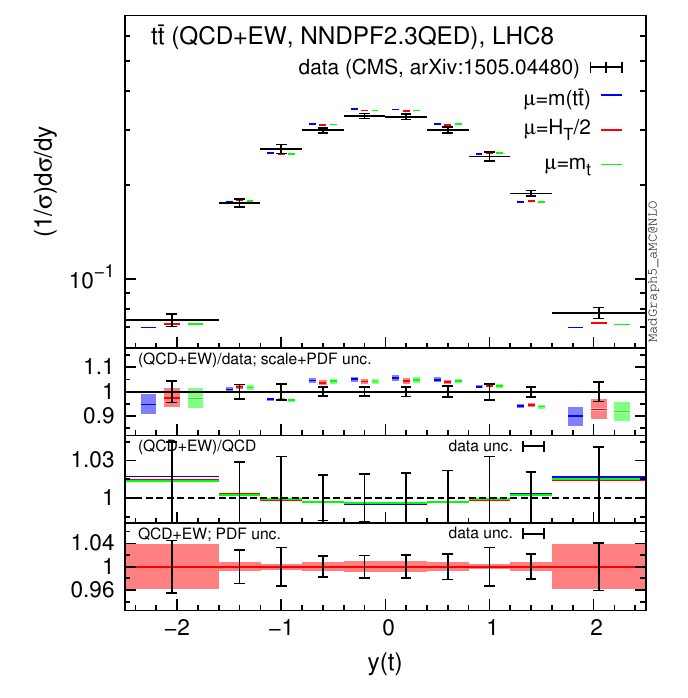}\\
\includegraphics[width=0.45\textwidth]{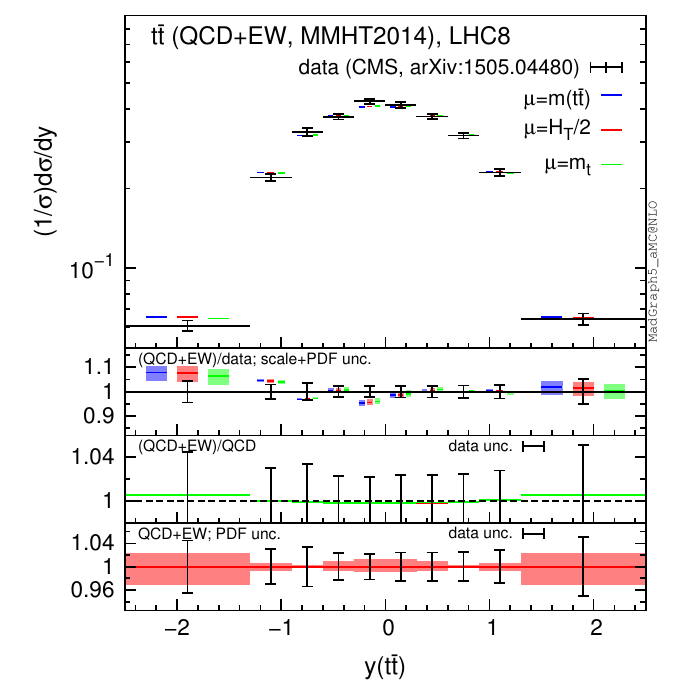}
\includegraphics[width=0.45\textwidth]{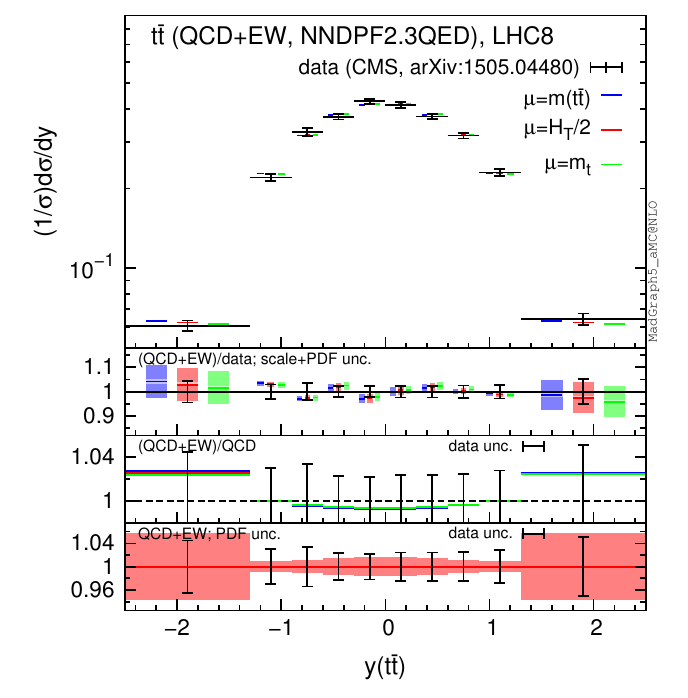}
\caption{Comparison of CMS data from ref.~\cite{Khachatryan:2015oqa} and NLO QCD+EW predictions computed with MMHT2104 (left) and NNPDF\-2.3\-QED (right) PDF sets.}
\label{fig:CMSdata}
\end{figure*}

For each differential measurement of ATLAS and CMS, we show two plots with the corresponding theoretical predictions obtained with the {\sc\small MMHT2014} PDF set~\cite{Harland-Lang:2014zoa} 
(plot on the left), and with {\sc\small NNPDF\-2.3\-QED}  (plots on the right), respectively.  At variance with what has been done in sect.~\ref{sec:res}, we do not set the photon PDF to zero in the left plots, but we directly use a PDF set that does not include neither LO QED contribution in the DGLAP running nor the photon PDF, as typically done in the experimental analyses and in QCD-only predictions. 
In the plots on the right we decided to use {\sc\small NNPDF\-2.3\-QED} in order to be conservative, since the effects due to the photon PDF are the largest for this PDF set. We explicitly verified that, as it could be expected from the discussion in sect.~\ref{sec:res}, using the  {\sc\small CT14QED} PDFs  or setting the photon PDF in {\sc\small NNPDF\-2.3\-QED} to zero, the relative corrections induced by $\Sigma_{\rm EW}$,  {\it e.g.} the $\Sigma_{\rm QCD+EW}/\Sigma_{\rm QCD}$ ratio, are the same as those obtained with {\sc\small MMHT2014}. Indeed the effect of the LO QED terms in the DGLAP running, which is present in {\sc\small CT14QED}  and {\sc\small NNPDF\-2.3\-QED} but not in  {\sc\small MMHT2014}, cancels in the ratio. On the other hand, this effect can have an impact on the cross section, especially on the $\Sigma_{\rm LO~QCD}$ term, and may be important for a direct comparison with data. This is another reason, besides the theoretical consistency, for which {\sc\small NNPDF\-2.3\-QED} or {\sc\small CT14QED} should be in general preferred when  NLO EW calculations are performed.

In each plot we present our predictions for a specific measurement, using three different definitions for the renormalisation and factorisation scales, {\it i.e.}, $\mu=m(\ttt)$ (blue), $\mu=H_T/2$ (as in sect.~\ref{sec:res}, red) and $\mu=m_t$ (green). In the main panel of each plot we present the $\Sigma_{\rm QCD+EW}$ predictions, without theory uncertainties, compared to data. In the first inset we display the ratio of   the $\Sigma_{\rm QCD+EW}$ over data,  including scale (light band) and PDF (dark band) uncertainties linearly added. 
In the second inset we show the central value of the ratio $\Sigma_{\rm QCD+EW}/\Sigma_{\rm QCD}$ for the three different scales and also the relative error in the measurement for each bin, with the error bar centred around  one. In the third inset we zoom the PDF uncertainty already included in the main panel. Specifically, we show the PDF uncertainty band of the $\Sigma_{\rm QCD+EW}$ predictions with $\mu=H_T/2$ normalised to their central value and we compare them again with the relative error in the measurement for each bin. Distributions corresponding to CMS analyses~\cite{Khachatryan:2015oqa} are shown in fig.~\ref{fig:CMSdata}, while  ATLAS distributions taken from ref.~\cite{Aad:2015hna} and ref.~\cite{Aad:2015mbv} are shown in fig.~\ref{fig:ATLASdata_m_y} and fig.~\ref{fig:ATLASdata_pt}, respectively.

Before discussing each plot, it is important to note that two different kinds of distributions are shown here, namely, differential distributions ($d\sigma$) and normalised distributions ($d\sigma/\sigma$).~\footnote{Consistently with what is done in the experimental analyses, we use as $\sigma$ the sum of the values of the bins of the distribution considered, without including overflows.} The $m(\ttt)$ distribution in fig.~\ref{fig:ATLASdata_m_y} and the boosted $p_T(t)$ distribution in fig.~\ref{fig:ATLASdata_pt} are  differential distributions, while all the other plots actually display normalised distributions, which exhibit much smaller scale and PDF uncertainties. Indeed, for normalised distributions both types of uncertainties are calculated directly for the $d\sigma/\sigma$ ratio, yielding  large cancellations between the numerator and the denominator.     
As we said, our aim is not to perform a direct comparison of data and SM predictions, which would need  QCD corrections beyond the NLO. Nevertheless,  it is interesting to note how the scale uncertainty shown in the main panel and  in the first inset of the plots, which reflects the typical QCD accuracy (NLO) of available Monte Carlo event generators, is of the  same order of the experimental precision for all the distributions considered here. Thus, although NNLO and NLO+NNLL have shown to reduce the dependence of the predictions on the renormalisation and factorisation scale, a few comments can be done about the scale choice for NLO simulations.  The plots for the $m(t \bar{t})$ distribution in  fig.~\ref{fig:ATLASdata_m_y} suggests that $m(t \bar{t})$ is not the natural scale for this process and indeed it artificially  leads to particularly large corrections also beyond NLO (see ref.~\cite{Pecjak:2016nee}). Similarly, the discrepancy between data and predictions observed in ref.~\cite{Czakon:2015owf} at NLO QCD for large $p_T(t)$, but not at NNLO QCD with fixed scale $\mu=m_t$, can be simply reduced at NLO using the scale $\mu=H_T/2$, as shown in the plots in the first line of fig.~\ref{fig:CMSdata}.

The most important information of the plots is however contained in the second and third insets. In the former we compare the experimental precision with the relative corrections induced by $\Sigma_{\rm EW}$ on the predictions at NLO QCD accuracy via the ratio $\Sigma_{\rm QCD+EW}/\Sigma_{\rm QCD}$.~\footnote{ In the case of normalised distributions, we actually display the ratio of $d\sigma/\sigma$ with both numerator and the denominator evaluated at QCD+EW accuracy over the same quantity at QCD accuracy $(d\sigma/\sigma)|_{\rm QCD+EW}/(d\sigma/\sigma)|_{\rm QCD}$, in order to correctly identify the effect of the electroweak corrections on the observable considered.} In the latter we compare  the experimental precision with the relative PDF uncertainties for the  $\Sigma_{\rm QCD+EW}$ predictions.  In the following, we discuss these comparisons for every measurement analysed here.

In the plots of the first line in fig.~\ref{fig:CMSdata}, we can see that the size of electroweak corrections is always smaller than the precision of the experimental measurement for $p_T(t)$ distributions.
However, it is reasonable to expect that in the Run-II at 13 TeV, with larger cross sections and higher luminosities, it will be possible to reach an experimental precision such that this effect could be probed, as it is also discussed in  ref.~\cite{Kuhn:2006vh} and supported by the first CMS analyses \cite{CMS-PAS-TOP-16-008,CMS-PAS-TOP-16-011} on 13 TeV data collected with only $2.3\ifb$ integrated luminosity. Nevertheless, it is important to note how results obtained with  {\sc\small MMHT2014} and  {\sc\small NNPDF\-2.3\-QED} are qualitatively different; the cancellation between photon-induced contributions and Sudakov logarithms at large $p_T(t)$ also strongly depends on the factorisation scale used. This is particularly important because the dependence of the $\Sigma_{\rm QCD+EW}/\Sigma_{\rm QCD}$ ratio on the PDF set and on the central scale choice is of the same order, or even exceeds, the scale uncertainty of the NNLO QCD predictions, which is at the percent level.~\footnote{The size of NNLO scale uncertainties  quoted in this section are  taken from ref.~\cite{Czakon:2015owf}. In particular the authors of ref.~\cite{Czakon:2015owf}  explicitly considered  the same  $p_T(t)$  and the $y(t)$ CMS distributions of fig.~\ref{fig:CMSdata} that are discussed in detail in the text.}
  The PDF uncertainties for $\Sigma_{\rm QCD+EW}$ predictions are smaller than the experimental precision both in the right and left plot, although  slightly larger values appear in the {\sc\small NNPDF\-2.3\-QED} case at large $p_T(t)$ due to the photon-induced contribution.

\begin{figure*}
\centering
\includegraphics[width=0.45\textwidth]{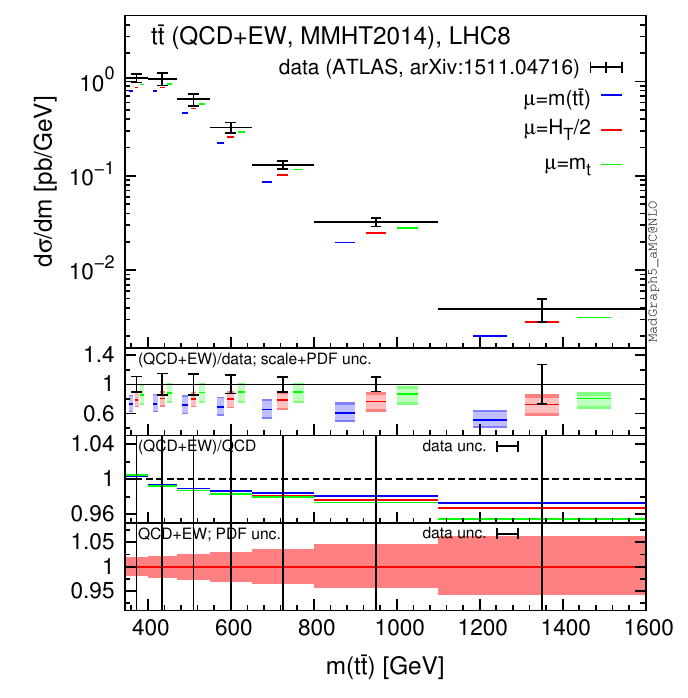}
\includegraphics[width=0.45\textwidth]{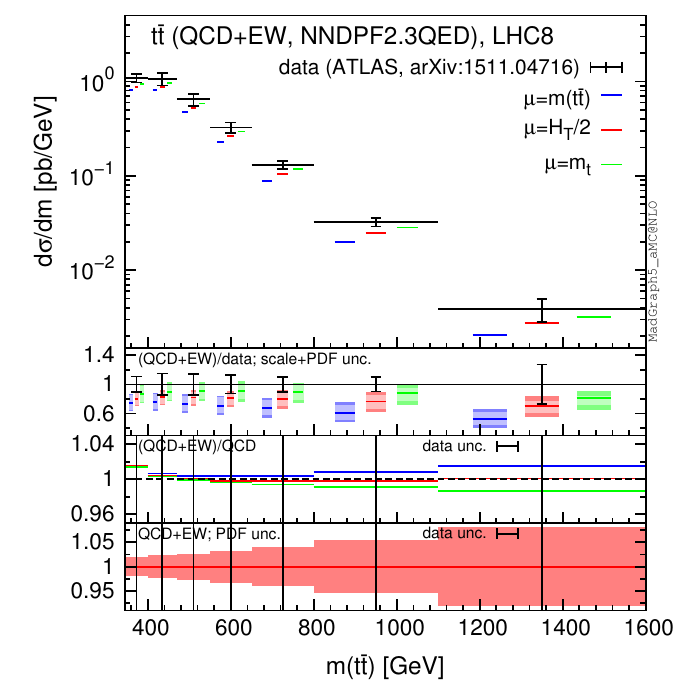}\\
\includegraphics[width=0.45\textwidth]{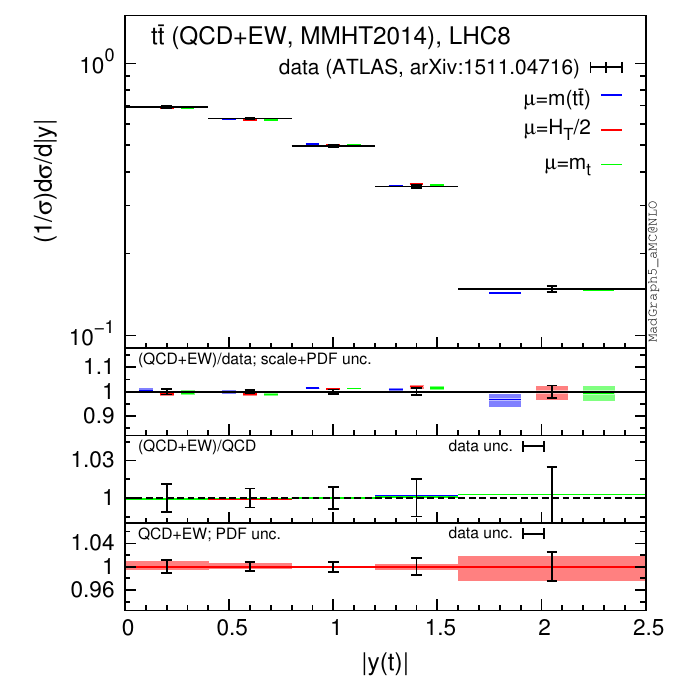}
\includegraphics[width=0.45\textwidth]{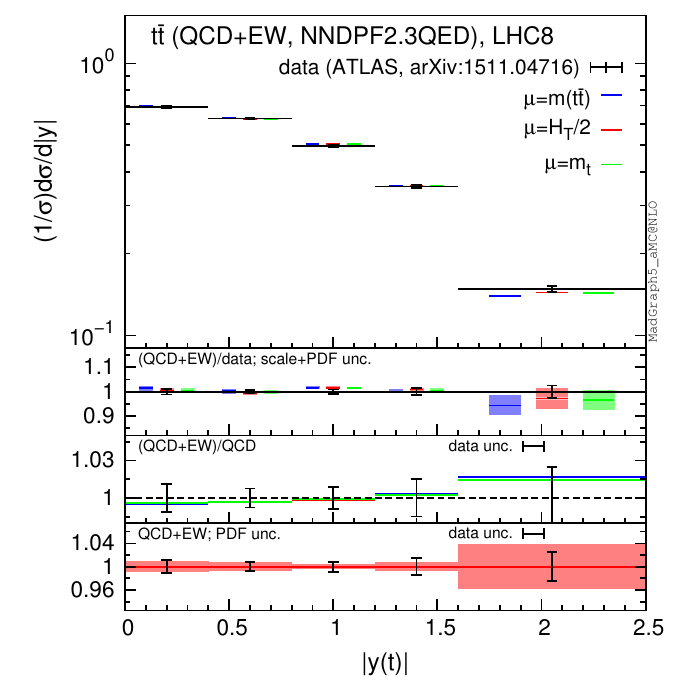}\\
\includegraphics[width=0.45\textwidth]{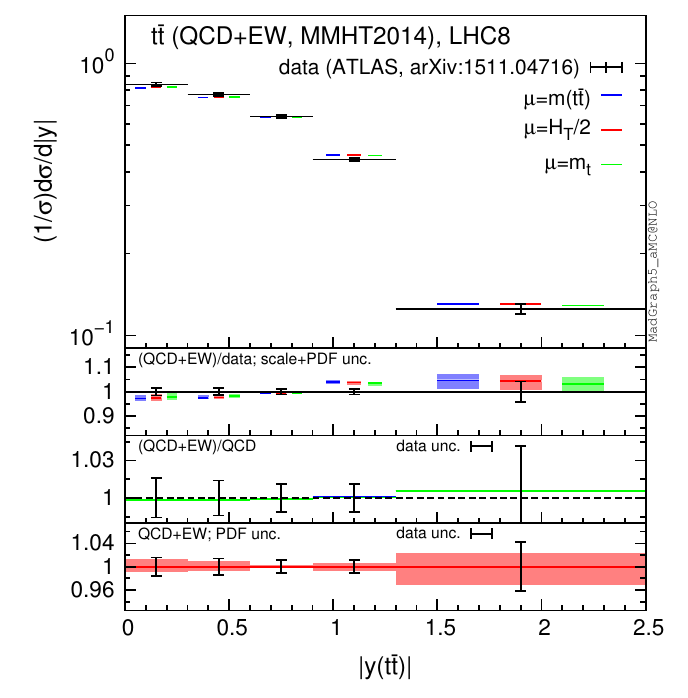}
\includegraphics[width=0.45\textwidth]{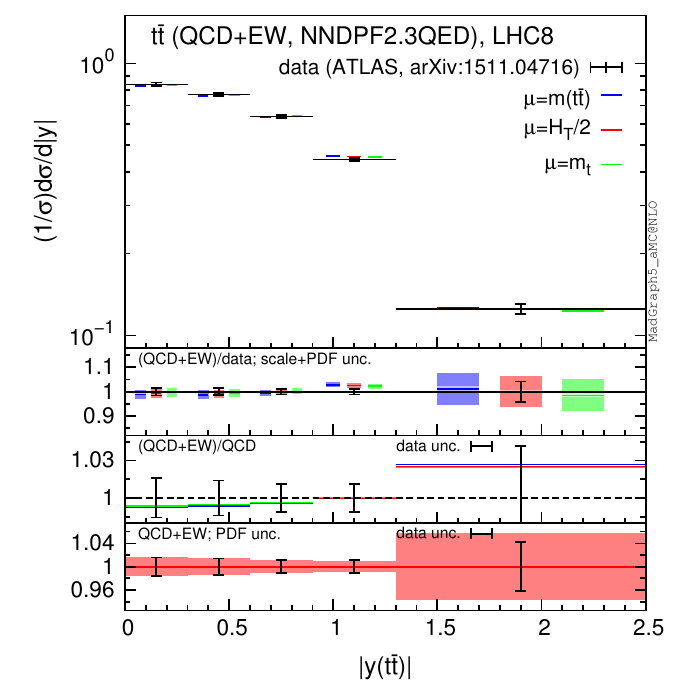}
\caption{Comparison of ATLAS data from ref.~\cite{Aad:2015mbv} and NLO QCD+EW predictions computed with MMHT2104 (left) and NNPDF\-2.3\-QED (right) PDF sets.}
\label{fig:ATLASdata_m_y}
\end{figure*}

\begin{figure*}
\centering
\includegraphics[width=0.45\textwidth]{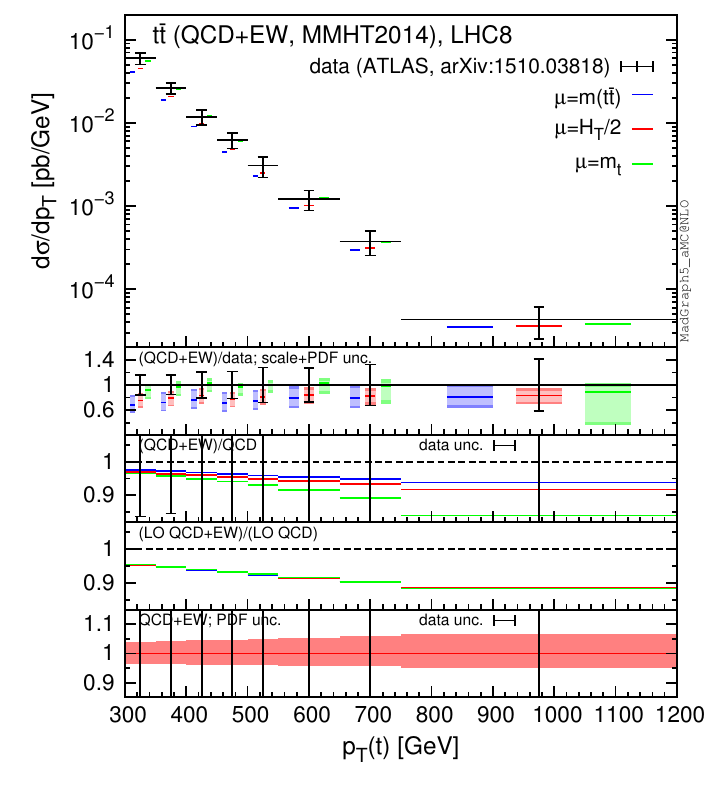}
\includegraphics[width=0.45\textwidth]{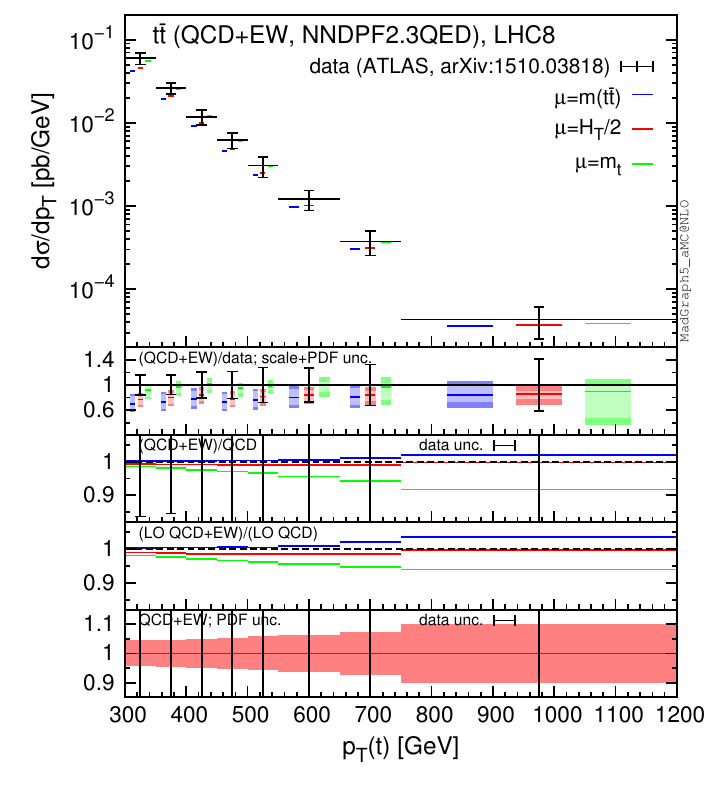}
\caption{Comparison of ATLAS data from ref.~\cite{Aad:2015hna} and NLO QCD+EW predictions computed with MMHT2104 (left) and NNPDF\-2.3\-QED (right) PDF sets.}
\label{fig:ATLASdata_pt}
\end{figure*}

The case of the $y(t)$ and $y(t \bar t)$ distributions in the plots in the second and third lines of both  figs.~\ref{fig:CMSdata} and \ref{fig:ATLASdata_m_y} clearly shows the relevance of EW corrections and photon-initiated processes for a trustworthy  comparison of theory predictions and data measurements at the LHC.  
  As can be seen in the first insets, scale uncertainties  are  smaller than the experimental errors and, for the $y(t)$ distribution, they are further reduced to  1\% level or even less by  NNLO QCD corrections, thus they are negligible. The theory uncertainties, as can be noted in the second insets, are completely dominated by PDF uncertainties. What is not negligible is the impact of the electroweak corrections. In the case of {\sc\small MMHT2014}, without photon-induced processes,  EW corrections are flat and thus their impact cancels in the $d\sigma/\sigma$ ratio. On the contrary, in the case of {\sc\small NNPDF\-2.3\-QED}, for large rapidities their impact is comparable to the experimental error, with a negligible dependence on the factorisation scale. Also, the enhancement in the peripheral region induces a small negative effect in the central region because of the fixed normalisation. PDF uncertainties for the $\Sigma_{\rm QCD+EW}$ are smaller or of the same order of the experimental precision in the case of  {\sc\small MMHT2014}, while they are larger in the case of {\sc\small NNPDF\-2.3\-QED} for large rapidities. All these effects are more prominent in the case of the ATLAS measurement (fig.~\ref{fig:ATLASdata_m_y}), where experimental errors are smaller. It is also worth to note that for rapidity distributions EW corrections do not receive Sudakov enhancements and, at this level of accuracy, besides the photon-induced contributions, {\it a priori} QED effects cannot be neglected.

The situation for the $m(t \bar{t})$ distribution in the first line of fig.~\ref{fig:ATLASdata_m_y} is very similar to the one for the $p_T(t)$ distribution, which has been already discussed for the CMS measurement (first line of  fig.~\ref{fig:CMSdata}) and it is also shown for the boosted regime in the ATLAS analysis in  fig.~\ref{fig:ATLASdata_pt}. However, these two {\it differential} distributions exhibit much larger scale uncertainties than the normalised $p_T(t)$ distribution in fig.~\ref{fig:CMSdata}. The inclusion of NNLO QCD corrections is crucial in order to pin down the scale uncertainties below the 5\% level, pointing  to the necessity of a combination of EW and NNLO QCD corrections for a reliable comparison of theory predictions and data measurements at the LHC. In all the $p_T(t)$ and $m(t \bar{t})$ distributions with the {\sc\small NNPDF\-2.3\-QED} set,  the dependence of the EW corrections on the factorisation scale is also particularly evident and in the following we describe the reason of this feature.

The dependence of the size (and possibly of the sign) of EW corrections on the scale choice can be traced to different effects. First of all, unlike QCD corrections, EW corrections to $q\bar{q}$ and $gg$ initial states do not involve neither $\alpha_s$ renormalisation nor the $\ord(\alpha_s)$ PDF counter terms that stabilise the $\mu_f$ dependence. Thus, without photon-initiated contributions,  both the $\mu_r$ and $\mu_f$ dependence 
of $\Sigma_{\rm LO~QCD}$ and $\Sigma_{\rm NLO~EW}$  is the same;  it is different for $\Sigma_{\rm QCD}$. This can be clearly seen in the left plot of fig.~\ref{fig:ATLASdata_pt}, where in addition to the $\Sigma_{\rm QCD+EW}/\Sigma_{\rm QCD}$ ratio in the third inset we included also the $(\Sigma_{\rm LO~QCD}+\Sigma_{\rm EW})/\Sigma_{\rm LO~QCD}$
ratio
~\footnote{
It is worth to note that without photon PDF this ratio corresponds to the NLO EW $K$-factor, $K^{\rm EW}\equiv (\Sigma_{\rm LO~QCD}+\Sigma_{\rm NLO~EW})/\Sigma_{\rm LO~QCD}$, which can be also seen as the ratio $\Sigma_{\rm QCD\times EW}/\Sigma_{\rm QCD}$, where  $\Sigma_{\rm QCD\times EW}\equiv\Sigma_{\rm LO~QCD}\times K^{\rm QCD} \times K^{\rm EW} $ is the multiplicative combination of results at NLO QCD and EW accuracy and we have defined as usual $K^{\rm QCD}=\Sigma_{\rm QCD}/\Sigma_{\rm LO~QCD}$. Thus the second and third insets in the left plot can be seen as a comparison of $\Sigma_{\rm QCD+ EW}$ and $\Sigma_{\rm QCD\times EW}$. On the other hand, the inclusion of $g\gamma$-channels in $\Sigma_{\rm LO~EW}$, such as in the right plot, invalidates the previous argument introducing a non-negligible contribution to $\Sigma_{\rm LO~EW}$.},
where the scale dependence  is the same in the numerator and the denominator. 
 On top of this, as anticipated in the last point in sect.~\ref{sec:PDF}, $gg$, $q\bar{q}$ and $g\gamma$ luminosities have a different dependence on $\mu_f$. Moreover, at LO the $g\gamma$ channel contributes only to $\Sigma_{\rm LO~EW}$, which features only one power of $\alphas$ and thus a different dependence on $\mu_r$ w.r.t. $\Sigma_{\rm LO~QCD}$.  While the Sudakov logarithms are negative and proportional to the $gg$ and $q\bar{q}$ contributions to  $\Sigma_{\rm LO~QCD}$, the $g \gamma $ contributions to $\Sigma_{\rm EW}$ are positive. Consequently, the cancellation of this two classes of contributions both entering $\Sigma_{\rm EW}$ depends on $\mu$. All these effects are present in any distribution on the right with {\sc\small NNPDF\-2.3\-QED}. However, in the rapidity distributions there is no Sudakov enhancement and $m_t\sim H_T/2$, especially for large rapidities, so only a small difference for the case $\mu=m(t\bar{t})$ is visible.

In conclusion, the impact of the photon PDF and also its dependence on the scale choice should be in general taken into account for the determination of the theory uncertainties.

\section{Conclusions and outlook}
\label{sec:concl}
In this work we studied the impact of EW corrections and pho\-ton-induced contributions on top-quark differential di\-stri\-bu\-tions at 8, 13 and 100 TeV. We compared predictions with two diffe\-rent modern PDF sets including the pho\-ton density and DGLAP evolution at NLO QCD + LO QED accuracy: the {\sc\small CT14QED} and {\sc\small NNPDF\-2.3\-QED} PDF sets. While contributions due to the photon PDF are negligible with  {\sc\small CT14QED}, this is not the case for {\sc\small NNPDF\-2.3\-QED}, where such contributions at 13 TeV are sizeable and are affected by large PDF uncertainties. At high $p_T(t)$ and $m(t\bar t)$, the photon-induced contributions can compensate the negative contributions of Sudakov logarithms or even change the sign of the EW corrections. Furthermore, we have shown that such a compensation strongly depends on the scale choice. In rapidity distributions, the impact of the $g\gamma$ initial state is sizeable in the peripheral region and much larger than NLO EW corrections, which do not receive large Sudakov enhancements in these kinematical configurations.

Increasing the energy of the collider, photon-induced channels become less relevant for a fixed value of $p_T(t)$ or $m(t \bar t)$, since smaller values of $x$ are probed and consequently the quark and gluon PDFs are much larger than the photon PDF. At 100 TeV, photon-induced channels are important only in the very boosted re\-gi\-me ($p_T(t)\gtrsim 5~\tev$ or $m(t \bar t)\gtrsim 10~\tev$), where Sudakov logarithms are negative and above the $20\%$ level.

For the same reason, at the LHC photon-induced contributions are relatively larger at  8 TeV than at 13 TeV. We computed their size for the same differential (and normalised) distributions already analysed by ATLAS and CMS, taking into account both experimental errors and theory uncertainties. Data from 8 TeV at large rapidities already appear to be  sensitive to the photon PDF;  with smaller experimental errors, as expected at 13 TeV, such a sensitivity may be reached also at large $p_T(t)$ and $m(t \bar t)$.

In conclusion, our analyses  showed two important points. First, differential observables in top-pair production, in particular $y(t)$ and $y(t \bar t)$ rapidities, can be used to improve the determination of the photon PDF within the {\sc\small NNPDF} approach. Second, given the status of the current available PDF sets, both EW corrections and photon-induced contribution  have to be taken into account for a correct determination of both central values and uncertainties of theoretical predictions. The first point is particularly relevant for a future fit with LHC 13 TeV data  at (N)NLO QCD as well NLO QED accuracy, which will presumably return a better determination of the photon PDF and  consequently, with smaller uncertainties, a more solid comparison with the ansatz for $\gamma(x,Q_0)$ used in  {\sc\small CT14QED}. The second point suggests also that {\it a priori} not only EW corrections but also photon-induced contributions may be in general important for processes at the LHC and must not be neglected. For the case of $\ttt$ production, given the experimental precision  that has been reached at the LHC and especially will be achieved in the Run-II, a combination of higher-order QCD corrections, beyond the NLO, and EW corrections, including photon-induced contributions, is mandatory for providing a precise and reliable SM predictions. This is relevant not only for top physics but also for all the analyses that feature $t \bar t$ production as background, notably those related to the Higgs boson and new physics.

\section*{Acknowledgments}

First and foremost, we want to thank Michelangelo Mangano for having encouraged us to perform this study and for his valuable comments. We also thank Valerio Bertone, Stefano Carrazza, Andrea Giammanco, Fabio Maltoni, Alexander Mitov and Carl Schmidt for interesting discussions and clarifications on different subjects treated in this paper. 
  
This work is done in the context of and supported in part (D.P.) by the ERC grant
291377, LHCtheory: Theoretical predictions and analyses of LHC
  physics: advancing the precision frontier.
I.T. is supported by the F.R.S.-FNRS ``Fonds de la Recherche Scientifique'' (Belgium) and in part by the Belgian Federal Science Policy Office through the Interuniversity Attraction Pole P7/37.
M.Z. is supported by the European Union's Horizon 2020 research and innovation programme under the Marie Sklodovska-Curie grant agreement No 660171 and in part by the ILP LABEX (ANR-10-LABX-63), in turn supported by French state funds managed by the ANR within the ``Investissements d'Avenir'' programme under reference ANR-11-IDEX-0004-02.

\bibliographystyle{apsrev4-1}
\nocite{apsrev41Control}
\bibliography{ttx_QCD_EW}

\end{document}